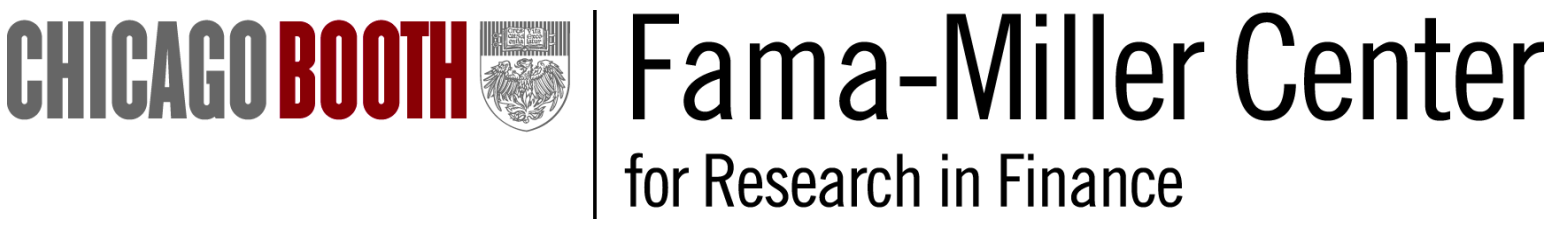

# ChatGPT and Corporate Policies

Manish Jha
Georgia State

Jialin Qian
Georgia State

Michael Weber
Chicago Booth School of Business and NBER

Baozhong Yang
Georgia State

# ChatGPT and Corporate Policies[*]

Manish Jha[†] *Georgia State*

Jialin Qian [‡] *Georgia State*

Michael Weber[§] *Chicago Booth and NBER*

Baozhong Yang[¶] *Georgia State*

**Abstract**

We create a firm-level ChatGPT investment score, based on conference calls, that measures managers' anticipated changes in capital expenditures. We validate the score with interpretable textual content and its strong correlation with CFO survey responses. The investment score predicts future capital expenditure for up to nine quarters, controlling for Tobin's $q$ and other determinants, implying the investment score provides incremental information about firms' future investment opportunities. Consistent with theoretical predictions, high-investment-score firms experience significant positive short-term returns upon disclosure, and negative long-run future abnormal returns. We demonstrate ChatGPT's applicability to measure other policies, such as dividends and employment.

[*]We are grateful to Hoa Briscoe-Tran, Will Cong, Zhi Da, John Graham, Robin Greenwood, John Griffin, Matt Grinblatt, Wei Jiang (discussant), Aakash Kalyani (discussant), William Mann (discussant), Gordon Phillips (discussant), Robert Reardon (discussant), Jinfei Sheng, Yuehua Tang (discussant), Stijn van Nieuwerburgh, Toni Whited, and Tengfei Zhang for helpful discussions and suggestions, and conference and seminar participants at the NBER Big Data and Securities Conference, the New York Federal Reserve FinTech Conference, the Q-Group Fall Seminar, the FIRS Conference, the CFEA Conference, the Australian Finance and Banking Conference, University of Texas at Dallas Finance Conference, the Boca-ECGI Corporate Governance Conference, the Crowell Prize Finalist Presentations at Panagora Asset Management, University of Delaware Weinberg/ECGI Corporate Governance Symposium, the JBF Generative AI Conference, the PHBS Finance Symposium, the Winter Institute of AISSEAI, International Monetary Fund, Luohan Academy, UC Irvine, University of Lausanne, Rutgers University, Georgia College, George Mason University, and Georgia State, for helpful comments and suggestions. We are thankful to John Graham for sharing the original Duke CFO Survey data.

[†]J. Mack Robinson College of Business, Georgia State University, Email: mjha@gsu.edu

[‡]J. Mack Robinson College of Business, Georgia State University, Email: jqian5@gsu.edu

[§]University of Chicago Booth School of Business and NBER, Email: michael.weber@chicagobooth.edu

[¶]J. Mack Robinson College of Business, Georgia State University, Email: bzyang@gsu.edu

# 1. Introduction

Understanding corporate policies is central to corporate finance. Investment policies, in particular, are key to corporate growth and aggregate fluctuations, with aggregate investment being the most volatile component of GDP (McConnell and Muscarella, 1985; Titman, Wei, and Xie, 2004; Bolton, Chen, and Wang, 2013). According to the neoclassical $q$-theory, Tobin's $q$ should be a sufficient statistic for describing firms' investment opportunities and policies (Hayashi, 1982; Peters and Taylor, 2017). Nonetheless, private information such as the expectations and plans of corporate managers may not yet be fully incorporated into market prices, even if the market is mostly efficient. Such information, in general, is not available for all firms, despite the availability and usefulness of information for a subset of firms provided by various surveys, e.g., the Duke University/Federal Reserve CFO Surveys and the Conference Board CEO Surveys.[1]

One way via which managers can convey their private information to market participants is through quarterly earnings conference calls that provide a wealth of information, including corporate managers' beliefs and expectations, to the public. Analyzing such information at a large scale is challenging because the length of a typical call is 8,000 words and thousands of companies report each quarter. Despite the progress in research tools in textual analysis in recent years, extracting complicated information such as the firms' expected investment policy has been beyond the reach of researchers, until the advent of the revolutionary AI tool, ChatGPT. Developed by Open AI, ChatGPT sets itself apart from previous AI models by being able to take long, sophisticated questions and provide detailed and sophisticated answers at the level of human experts.

In this study, we use ChatGPT to extract firm-level corporate expectations of future investment policies and aim to answer the following research questions: Can an advanced AI model such as ChatGPT help understand corporate policies? Does the ChatGPT-extracted expected investment policy provide information beyond existing measures of investment opportunities, such as Tobin's $q$ or cash flows? Does such information have further implications on asset prices and returns? We address these questions using 74,429 conference call transcripts for 3,863 unique companies from 2006 to 2020. We provide conference call transcripts with questions about the expected future capital expenditures to the ChatGPT model to retrieve quantitative assessments of future increases and decreases in investment and construct a *ChatGPT Investment Score*.

[1] More information on these data is available at https://www.richmondfed.org/cfosurvey/ and https://www.conference-board.org/topics/CEO-Confidence.

We adopt several methods to validate this measure. First, since the Duke CFO surveys directly ask managers to answer questions regarding their future investment plans, we compare our investment score with the Duke survey responses from CFOs of the same company. We find a strong positive correlation between our measure and the survey answers on firms' expected capital expenditure policy. Second, the time series of the average investment score in our sample and the average future changes in capital expenditure exhibit similar trends over time and align well with each other. Third, we examine the time variation in the industry-level average investment scores and identify patterns consistent with major changes in the economy, e.g., the software and biotech industries expect an increase in investment during the Covid pandemic, in contrast to other industries that substantially cut expected investment. Finally, we ask ChatGPT to provide excerpts from conference call transcripts to support its assessment of the highest and lowest investment scores. The responses from ChatGPT reveal key phrases and sentences that are clearly interpretable by humans. This latter validation provides an important advantage of ChatGPT over some previous AI models – the interpretability of its outputs, which lends credence to the generated investment score.

To motivate the empirical analyses of the paper, we build a stylized model of corporate disclosure and investment. In the model, a corporate manager discloses a private signal that contains information about future investment opportunities. The model makes several predictions. First, future corporate investment is increasing in the disclosed signal. This prediction is consistent with the intuition that optimal investment is aligned with Tobin's $q$ (Hayashi, 1982; Abel and Eberly, 1994; Peters and Taylor, 2017). Second, the short-term return of the firm around the disclosure date is *positively* associated with the signal. Intuitively, an improved investment opportunity set allows the firm to adopt investment decisions that ultimately increase the value of the firm. Finally, the future expected return of the firm is *negatively* associated with the managerial signal or expectation of future investments. This prediction is closely related to the prediction of investment-based asset pricing theories (e.g., Liu, Whited, and Zhang, 2009) that high-investment firms generate lower future returns than low-investment stocks. Such an investment factor indeed features prominently in new standard asset pricing models, e.g., the Fama-French 5-factor model (Fama and French, 2015) and the $q$5-factor model (Hou, Mo, Xue, and Zhang, 2021).

To the extent the ChatGPT-based investment score represents firms' investment expectations that are not yet fully incorporated in market prices, we would expect that it contains incremental predictive power for future capital expenditure relative to Tobin's $q$ as motivated by the neoclassical $q$-theory and its extension, total $q$, that incorporates intangible capital (Peters and Taylor, 2017). We find the ChatGPT investment score bears a significant and positive relationship with future investment, keeping constant other determinants of firm investment. A one-standard-deviation increase in the investment score is associated with a 0.055 standard-deviation increase in capital expenditure in the quarter after the conference call. The economic magnitude is meaningful and sizeable and corresponds to about two-thirds of the sensitivity of capital expenditure to total $q$. This relation is robust to controlling for total $q$, total cash flows, lagged capital expenditure, other firm characteristics, as well as firm and time fixed effects, suggesting the investment score indeed contains new, incremental information derived from managerial private information and expectations.

The significant predictive power of the investment score for future investment continues to hold for the subsequent nine quarters, which partially indicates the long-term nature of managers' expectations but likely also reflects the long-term nature of large investment projects. The cumulative increase in future investment over the next nine quarters due to a one-standard-deviation increase in the investment score is 0.51% of book assets, or 32% of a standard deviation of quarterly capital expenditure. Furthermore, the ChatGPT investment score contains information beyond future physical investment and can help predict other forms of investment, including intangible investment, Research and Development (R&D), and total investment in both the short term and the long run.

To alleviate the concern that Tobin's $q$ may be measured with error, which can contribute to the predictive power of the ChatGPT investment score, we employ the cumulant regression approach proposed by Erickson, Jiang, and Whited (2014), which corrects for measurement bias in $q$. The results show the ChatGPT investment score continues to provide significant additional predictive power to $q$ measures even after this bias-correction of the coefficient estimates.

Given that the ChatGPT investment score captures new information regarding firms' future investment opportunities and complements the information in current investment and Tobin's $q$, our model predicts the ChatGPT investment score should also be negatively related to future stock returns. Our tests confirm this hypothesis. The ChatGPT investment score is significantly and negatively associated with raw returns and factor-adjusted abnormal returns over the following quarter, controlling for total $q$ and past returns. A one-standard-deviation increase in the investment score corresponds to $-1.83\%$, $-1.49\%$, and $-1.42\%$ in raw returns, FF5-adjusted returns, and $q5$-adjusted returns in the quarter subsequent to the earnings call, respectively. Similar to investment, the return predictability also persists for up to nine quarters after the earnings call. The existence of such abnormal returns suggests that the market does not fully incorporate information already contained in public corporate earnings calls, and an advanced AI model like ChatGPT is able to extract such information efficiently. Employing such AI models can, thus, help investors extract useful information and potentially make the market more efficient.

We also find that the ChatGPT investment score is significantly and positively associated with the announcement returns around the conference call date, even after controlling earnings surprises and the general textual sentiment of the call transcripts, consistent with another model prediction. This evidence further confirms the information value of the AI-generated managerial expectation variable.

Next, we study cross-sectional heterogeneity in the association between the ChatGPT-based investment score and future investments. Managerial expectations and forecasts are likely to be more informative and valuable for more opaque firms and firms operating in a more dynamic and uncertain environment. We proxy for the nature of the environment a firm operates in by industry competition, firm size, and stages of the product life cycle (Hoberg and Maksimovic, 2022). The predictive power of the ChatGPT investment score is particularly pronounced for firms that are smaller, are in their initial stages of the product lifecycle, and are operating in more competitive landscapes, consistent with the above hypothesis.

We further explore the factors driving firms' expected investment based on ChatGPT investment scores by categorizing investment explanations into three main groups: Operational Efficiency and Cost Control, Strategic Planning and Business Focus, and Regulatory Compliance and Market Conditions. Temporal analysis reveals that during economic downturns, firms predominantly cite market conditions and operational efficiency as investment drivers, while strategic planning dominates during stable economic periods. All three categories predict long-term capital expenditure, with regulatory compliance and market conditions showing a shorter predictive horizon.

ChatGPT might use information beyond the corpus of the conference call transcript and especially future information in generating the investment score and hence, researchers and market participants might not be able to use the information in real time to make investment decisions. We provide two additional tests to alleviate these concerns. The first test is an out-of-sample test, in which we rerun our main analysis for the period of 2021Q4 to 2022Q4, after the end of the training period of ChatGPT 3.5 in September 2021. In the second test, we mask all dates, firm, person, and product names from the conference calls and redo the ChatGPT score from the masked transcripts. Our main results continue to hold in both tests. We also investigate the predictive power of other large language models, such as RoBERTa. While they are also capable of understanding corporate policies, ChatGPT exhibits superior performance in a horse race. Furthermore, we perform a number of additional sensitivity checks that incorporate additional control variables and alternative definitions of the ChatGPT score. The results are robust.

Despite the focus of this study on corporate investment policies, we also investigate whether our methodology can be applied to other corporate policies. In particular, we employ ChatGPT in a similar way to obtain managerial expectations of changes in dividend payment and employ-ment policies and construct ChatGPT-based dividend and employment scores. These AI-based expected policy measures are strongly correlated with the expected policies reported in the Duke CFO Survey responses for the same set of firms. Therefore, our approach has the potential to be applicable to a wide range of corporate policies.

This paper makes several contributions. First, it is the first paper to apply the cutting-edge AI tool, ChatGPT, to extract managerial expectations of corporate policies from corporate earnings calls and validate the AI-based policy measures empirically. Our methodology can be applied to a broad range of policies and expectations. Second, the ChatGPT investment measure provides a new, real-time measure of expected investment that complements the $q$ measures in classical and extended $q$-theories. Third, our method can be used to expand and complement existing surveys of executives, which can be especially helpful given the decline in survey response rates in the US in the past decade, especially after the Covid pandemic (Pickert, 2023) and given the high costs of running surveys of firms (Weber et al., 2022). Fourth, AI interpretability is an important concern, given the increasing prevalence of AI in financial and economic studies and the challenge of explaining certain "black box" models. Our approach allows for an interpretable application of AI, as humans can read and understand the arguments ChatGPT gives when making decisions.

We contribute to several streams of literature. First, our paper is related to the literature on the investment-$q$ relation. Despite theories that establish strong links between Tobin's $q$ and investment (Tobin, 1969; Hayashi, 1982; Abel and Eberly, 1994), their empirical relation had been weak.[2] A large literature explains this puzzling discrepancy. For example, Erickson and Whited (2000, 2012) and Erickson, Jiang, and Whited (2014) propose several approaches, including the GMM and cumulant regression methods, to remove measurement errors in $q$. Philippon (2009) uses bond prices to obtain a more accurate measure of $q$. Peters and Taylor (2017) introduce a novel measure of total $q$, which includes a measure of intangible capital, and find that total $q$ significantly improves the investment-$q$ relation.[3] Andrei, Mann, and Moyen

[2]See surveys by Hassett and Hubbard (1997) and Caballero (1999).

[3]See also the literature that develops various measures of intangible capital, e.g., Corrado and Hulten (2010, 2014), Eisfeldt and Papanikolaou (2013, 2014), Falato, Kadyrzhanova, Sim, and Steri (2022), and Ewens, Peters, and Wang (2019).

(2019) find the investment-$q$ relation is stronger in recent years because future cash flows and $q$ are both driven by innovation and learning. Our AI-based investment score provides new information for firms' future investment opportunities that complements Tobin's $q$ and total $q$, which can help researchers and regulators to better understand corporate investment and its consequences for the economy.

Second, our paper pertains to the feedback literature, in which managers learn from prices when making investments and other corporate decisions (Chen, Goldstein, and Jiang, 2007; Bakke and Whited, 2010; see the surveys Bond, Edmans, and Goldstein, 2012 and Goldstein, 2023 for comprehensive discussions of this literature). Our findings suggest that the other direction of the link is also important: the market can also learn from managers. Information extracted from corporate disclosure, such as expected corporate policies, can provide important new insights to investors and the market.

Third, our study relates to the literature on managerial and firm expectations. Surveys have been a powerful tool for researchers to obtain access to information that is not available in standard datasets. They are particularly instrumental in obtaining information regarding agents' beliefs and expectations (e.g., D'Acunto, Hoang, and Weber, 2022; Coibion, Gorodnichenko, and Weber, 2022; Weber et al., 2022), studying how they relate to corporate policies (e.g., Coibion, Gorodnichenko, and Kumar, 2018; Coibion, Gorodnichenko, and Ropele, 2020; Candia et al., 2023), or shedding light on corporate operations and decision-making processes (e.g, Graham and Harvey, 2001; Graham, Harvey, and Puri, 2013; Graham, Grennan, Harvey, and Rajgopal, 2022). Our approach can complement existing surveys, generate measures based on executives' plans and discussions for a large sample of firms, and provide a new set of tools and data for researchers.

Finally, our approach provides a step forward for textual analysis. Researchers have utilized textual analysis to analyze unstructured text information such as the levels and extent of sentiment (e.g., Tetlock, 2007; Hanley and Hoberg, 2010; Loughran and McDonald, 2011; Jiang, Lee, Martin, and Zhou, 2019; Jha, Liu, and Manela, 2021), political risk (Hassan, Hollander, Van Lent, and Tahoun, 2019), cyber risk (Florackis et al., 2023), synergies in M&As (Hoberg and Phillips, 2010), business news and corporate filings topics (Cong, Liang, Zhang, and Zhu, 2019; Bybee, Kelly, Manela, and Xiu, 2023) or corporate culture (Li, Mai, Shen, and Yan, 2021).[4] Other large language models such as BERT have been increasingly applied in various studies, su corporate disclosure policies (Cao, Jiang, Yang, and Zhang, 2023), sentiment toward finance (Jha, Liu, and Manela, 2022), patent's exposure to changes in patent law (Acikalin, Caskurlu, Hoberg, and Phillips, 2022), or news and expected returns (Chen, Kelly, and Xiu, 2023), among others. Recently, researchers have started to use ChatGPT to analyze sentiments of news headlines (Lopez-Lira and Tang, 2023), parse Federal Reserve announcements (Hansen and Kazinnik, 2023), examine redundant discussions in conference calls (Kim, Muhn, and Nikolaev, 2023), and forecast innovation success (Yang, 2023). We show ChatGPT can help extract information about complex concepts such as future corporate policies. Furthermore, such information is interpretable, which can increase AI's use in facilitating the decision-making of humans and help achieve synergies between man and machine (e.g., Armour, Parnham, and Sako, 2022; Cao, Jiang, Wang, and Yang, 2022; Brogaard, Ringgenberg, and Roesch, 2023).

## 2. Model

In this section, we construct a model of investment and disclosure to provide testable predictions for managerial expectations of future investment opportunities and guide our empirical analyses. We build a stylized model given the qualitative nature of our predictions.

We consider a three-period model with dates running from $t$ to $t+2$. A firm $i$ is endowed with $K_t$ units of productive capital at time $t$.[5] In period $s$ for $s \in \{t, t+1\}$, the firm makes investment decision $I_s$ with investment and capital adjustment cost according to

$$c(I_s, K_s) = c_1 I_s + c_2 K_s \left(\frac{I_s}{K_s}\right)^2. \tag{1}$$

Capital depreciates at rate $\delta$, and thus $K_{s+1} = (1-\delta)K_s + I_s$. The profit function $\pi_s = \pi_s(K_s)$ has constant return to scale. The firm pays dividends $d_s$ to investors after investment is made.

[4] See Loughran and McDonald (2016) for a comprehensive review of the use of textual analysis in accounting and finance.

[5] For simplicity, we do not distinguish physical and intangible capital in the model. The qualipredictions hold in an extension of the model that allows both types of capital, similar to Peters and Taylor (2017).

Let $M_s$ be the stochastic discount factor that prices asset returns from time $s$ to $s+1$. The firm's manager selects the optimal policies that maximize shareholders' value in that period. In the final period $t+2$, the firm produces profits and does not make further investment as there are no future returns. The terminal market value of the firm is thus $V_{t+2} = \pi_{t+2}(K_{t+2})$. The market value of the firm in period $t+1$ (after investment and dividend decisions) is given by

$$V_{t+1} = E_{t+1}[M_{t+1}\pi_{t+2}(K_{t+2})]. \tag{2}$$

and the expected return in period $t+1$ is

$$E_{t+1}[R_{t+1}] = E_{t+1}\left[\frac{V_{t+2}}{V_{t+1}}\right]. \tag{3}$$

Tobin's $q_{t+1}$, which represents the future investment opportunities, is given by $q_{t+1} = \frac{V_{t+1}}{K_{t+2}}$.[6] We assume that the Tobin's $q$ consists of several components: $q_{t+1} = q_t^e + q_t^m + \epsilon_{t+1}$, where $q_t^e$ is the expected value of Tobin's $q$, $q_t^m$ is a private signal from the manager, and $\epsilon_{t+1}$ is a random component not known to investors or the manager.

In the model, the manager publicly discloses her signal $q_t^m$ at the beginning of period $t+1$, before investment and dividend decisions are made, which allows the market to update its expectation of $q_{t+1}$ from $q_t^e$ to $q_t^e + q_t^m$. In our later analyses, we use ChatGPT to extract the managerial expectations about future investment opportunities, which then serve as a proxy for $q_t^m$.

Let $V_{t+1,d-}$ and $V_{t+1,d+}$ be the market value of the firm right before and after the disclosure date of the firm. Note that both of these dates are before investment and dividend decisions are made in period $t+1$. We have the following propositions that provide predictions that link the managerial expectation signal $q_t^m$ for firms' future investment decisions and short-term and long-term returns. We relegate proofs to Appendix D.

[6] Note the market value is computed after investment the decision $I_{t+1}$ is made and therefore the capital quantity $K_{t+2} = (1-\delta)K_{t+1} + I_{t+1}$ is known at that time.

**Proposition 1.** *Future investment* $I_{t+1}$ *is increasing in the managerial signal* $q_t^m$.

This proposition reflects the classical result of the neoclassical models (e.g., Hayashi, 1982; Abel and Eberly, 1994) that average $q$ is equal to the marginal $q$, which determines investment when capital adjustment is costly.

**Proposition 2.** *When the manager reveals the private signal* $q_t^m$ *about investment opportunities, other things equal, the short-term return* $\frac{V_{t+1,d+}}{V_{t+1,d-}}$ *is increasing in* $q_t^m$.

The intuition of this proposition is that the increased investment opportunity is a positive signal for firm value and thus leads to a positive short-term return upon corporate disclosure.

**Proposition 3.** *Other things equal, the long-term expected return* $E_{t+1}[R_{t+1}]$ *is decreasing in* $q_t^m$.

The intuition of this result is that other things equal, high-investment firms have higher current market values and thus lower expected future returns, analogous to the investment-based asset pricing theories (e.g., Liu, Whited, and Zhang, 2009).

## 3. Data

In this section, we discuss the different datasets we use as well as the variable construction.

### 3.1. Data Sources and Sample

We rely on several data sources. First, we use public companies' conference call transcripts as our primary text source for the purpose of extracting firms' outlooks on corporate policies. Second, we obtain the quarterly Duke CFO survey firm-level data which has been initiated and analyzed in Graham and Harvey (2001).[7] Third, we utilize Compustat and CRSP to obtain corporate accounting variables and stock returns.

The primary text dataset we use encompasses earnings call transcripts from 2006 to 2020, sourced from the Seeking Alpha website.[8] These transcripts are compiled from quarterly earnings calls conducted by senior executives, such as CEOs and CFOs, during which they provide investors and analysts with a comprehensive overview of their firm's overall performance. Along with discussing their company's quarterly performance, executives often provide forward-looking statements and their own assessments of the business. Managers also share their business strategies and operational plans with investors. Furthermore, during the conference calls, analysts and potential investors can pose questions to the management and further explore different aspects of the firms' operations, plans, and performance.

[7] We are grateful to John Graham for sharing the data from CFO surveys.

[8] Available at https://seekingalpha.com/earnings/earnings-call-transcripts.

We sample a total of 160,195 earnings call transcripts spanning the years 2006 to 2020. We first merge the earnings call transcripts with CRSP and Compustat by using the stock ticker, the title and date of the earnings calls. This step reduces the sample size to 115,620 transcripts. We then obtain financial and balance sheet variables from Compustat, and stock returns from CRSP. After requiring all main variables in our analyses to be non-missing, the final sample consists of 74,429 firm-quarter-level conference calls and merged corporate data from 2006 to 2020, representing 3,878 unique US public firms.

### 3.2. Variables

Our first measure of investment is *Capital Expenditure*, which is the capital expenditure scaled by total book assets. We also define several variables following Peters and Taylor (2017): *Intangible Capital*, calculated from accumulating R&D and a proportion of Selling, General, and Administrative (SG&A) expenses; *Physical Capital*, PP&E; *Total Capital*, the sum of *Intangible Capital* and *Physical Capital*; *Total q*, the ratio of market capitalization to *Total Capital*; *Physical Investment*, which is capital expenditure scaled by *Total Capital*; *Intangible Investment*, which is R&D + 0.3 × SG&A expenses, scaled by *Total Capital*; *Total investment*, the sum of *Physical Investment* and *Intangible Investment*. We introduce the ChatGPT-predicted capital expenditure plan, *ChatGPT Investment Score* in Section 4.1.

We include the following control variables in our analyses: *Size*, the natural logarithm of total book assets at the end of the quarter; *Total Cash Flow*, as described in Peters and Taylor (2017), the ratio of *Total Capital* to the sum of income before extraordinary items, depreciation expenses, and after-tax *Intangible Investment*; and *Leverage*, the book value of debt divided by total book assets at the end of the quarter. We provide the definitions of all variables in Appendix A.

### 3.3. Duke CFO Survey

The Duke CFO survey is a comprehensive survey of managerial outlooks of the economy, firm performance, and corporate policies. The survey was initiated by Graham and Harvey (2001) and continued at a quarterly frequency by the Fuqua Business School at Duke University until 2020Q1, after which it is jointly run by Duke and the Federal Reserve Banks of Richmond and Atlanta.[9] We focus on the following survey question:

> "Relative to the previous 12 months, what will be your company's PERCENTAGE CHANGE during the next 12 months? _________% *[Corporate Policy]*"

In the prompt, *[Corporate Policy]* can refer to a number of corporate policies, including "Capital Spending," "Number of domestic full-time employees," etc. We gather firms' responses to this question on "Capital Spending" and create a variable *CFO Survey Investment* at the firm-quarter level.

We match firms in the Duke CFO Survey data to the conference call data using multiple identifiers, including Compustat's global value keys (GVKEY), CRSP's permanent company number (PERMNO), and the unique respondent id in the Duke Survey. In total, we are able to match 1,707 surveys to their corresponding conference calls. Since not all respondents provided answers to every survey question, the sample sizes vary for different questions.

## 4. Empirical Methodology

We now discuss the construction of the ChatGPT investment score, provide summary statistics, and validate the measure.

[9]The survey questions and summary results are available at `https://cfosurvey.fuqua.duke.edu/` and `https://www.richmondfed.org/research/national_economy/cfo_survey`.

### 4.1. ChatGPT Investment Score

ChatGPT is an artificial intelligence chatbot developed by OpenAI based on the company's Generative Pre-trained Transformer (GPT) series of large language models. The GPT architecture is based on transformers, which are deep learning models designed to handle sequential data, such as natural language texts. Transformers consist of multiple layers of self-attention mechanisms that allow the model to capture dependencies between words in a sentence. Google's BERT (Bidirectional Encoder Representations from Transformers), released in 2018, is the first transformer-based large language model with acclaimed success. Another milestone is the very large GPT-3 model, trained on 45TB of data and with 175 billion parameters, released by OpenAI in June 2020. ChatGPT, launched on November 30, 2022, took the world by surprise with its capability of offering detailed and articulate responses spanning various domains of knowledge.

One way to understand ChatGPT would be to think of it as a giant robot that has read millions of books, papers, and articles, and learned the textual patterns therein. When prompted with a question, ChatGPT uses what it has learned (through a combination of supervised and reinforcement learning techniques) to understand the meaning and produce (predict) the best responses.

We prefer using ChatGPT over human reading for conference call text analysis for several reasons. First, ChatGPT provides consistent evaluations because it doesn't rely on other contemporaneous information given an approprioate prompt or personal opinions that could introduce biases, ensuring a reliable, consistent, and objective assessment of conference call content. Second, conference calls can be lengthy, often exceeding seven thousand words, making it challenging for humans to consistently provide accurate responses for reading comprehension tasks. Third, as an algorithm, ChatGPT does not have the capacity constraints of humans and can process a large number of texts in a short time frame.

In addition, when compared to other machine learning models such as BERT, ChatGPT is particularly well-suited for analyzing conference calls. Its training in a conversational context enables a better understanding of texts presented in a dialogue format. ChatGPT can effectively maintain context and coherence throughout the conversation, which proves beneficial for han-dling the interactive nature of back-and-forth exchanges commonly observed during earnings conference calls.

We use ChatGPT 3.5 as the large language model to process texts.[10] ChatGPT has a total limit of 4,096 tokens or around 3,000 words for input and output combined. Therefore, we first split each conference call into several chunks of length less than 2,500 words to conserve sufficient space for output. A typical earnings call is composed of three chunks or parts. To obtain the firms' expected capital expenditure from the earnings call transcripts, we provide the following prompt to ChatGPT.

> The following text is an excerpt from a company's earnings call transcripts. You are a finance expert. Based on this text only, please answer the following question. How does the firm plan to change its capital spending over the next year? There are five choices: Increase substantially, increase, no change, decrease, and decrease substantially. Please select one of the above five choices for each question and provide a one-sentence explanation of your choice for each question. The format for the answer to each question should be "choice - explanation." If no relevant information is provided related to the question, answer "no information is provided."
>
> *[Part of an earnings call transcript.]*

We extract the choice from the response of the model for each chunk of the earnings call and then assign a score of -1, -0.5, 0, 0.5, and 1 for each of the given choices (Decrease substantially; Decrease; No change; Increase; Increase substantially), respectively. If ChatGPT generates an answer "no information is provided," we assign a value of zero to the score. A potential drawback of ChatGPT is its occasional tendency to confidently provide inaccurate information. To combat inaccurate results, we ask ChatGPT to provide an explanation for each answer. We manually read and check the choices and explanations given by ChatGPT for a random sample of conference calls and find the mismatch rate of choice-explanation to be less than 1%, indicating a high level of accuracy. Therefore, we do not make any adjustments to the generated choices and assigned scores. We then take the average of the scores across multiple chunks of one earnings call to obtain a firm-quarter-level measure, *ChatGPT Investment Score*. Our main results are robust to alternative ways of aggregating text-chunk scores (see Section 6).

[10] The most recent version of ChatGPT based on GPT 4.0 is still prohibitively expensive at the time of the writing of this version of the paper for analyzing the entire conference call corpus.

To understand how ChatGPT is able to infer future investment policies from the conference call transcripts, we construct word clouds for paragraphs with high or low ChatGPT predicted scores (1 and $-1$). Specifically, we first extract all chunks of conference calls to which ChatGPT assigns an investment score of $-1$ or 1, respectively. We then ask ChatGPT to provide a one-sentence explanation of the reason for assigning such a score. Based on the answers from ChatGPT, we compile the word clouds of bi-grams for the high and low-investment-score groups and display them in Figure 1.

[Insert Figure 1 Here]

The word clouds reveal distinct themes. Both word clouds contain certain common bigrams that are associated with capital expenses, such as: “cash flow," “capital spend," “capital expenditure," etc. In the word cloud for low-investment-score texts (subfigure (a)), we see bigrams such as “cost reduction," “significantly reduce," “substantially reduce," “reduce cost," etc., indicating management’s plans to reduce capital expenditure. On the other hand, subfigure (b) for high-investment-score texts shows bigrams such as “revenue grow," “revenue growth," “term growth," “growth opportunity," etc., suggesting management’s willingness to invest in growth. We also provide several example text excerpts from conference call texts with high and low investment scores in Appendix B. The examples demonstrate similar topics as the word clouds but offer more detailed reasoning, e.g., “accelerate our investments in Safety Products, Intelligrated and other growth opportunities,” and “the optimization plan includes some business and international market exits, all of which had negligible margin.”

To create a visual representation of the changes in *ChatGPT Investment Score* over time, we compose an aggregate ChatGPT investment score by taking the cross-sectional average across all firms for each quarter in our sample. We then plot the time-series of this aggregate ChatGPT investment score with that of the average change in capital expenditure in Figure 2.

The trends in the two time-series are very similar over our entire sample period. Note that we focus on the trends, not the specific levels, since the two investment variables are constructed using completely different approaches. Furthermore, the aggregate ChatGPT investment score correctly identifies the 2007-2009 Great financial crisis and the 2020 Covid-19 crisis, as well as the investment booms following these episodes. The evidence from the figure indicates that our ChatGPT-based measure captures what it is intended to measure, that is, firms' expected investment.

[Insert Figure 2 Here]

In Figure 3, we drill down to the industry level and show the yearly trend across major sectors. Again, the ChatGPT investment score identifies plausible trends in industry investment across economic cycles. The ChatGPT-generated score captures the hardest-hit industries in the two crises: the Retail/Wholesale sector in 2007-2009, and the Transport/Energy sector in 2020. It also captures the resilience of the Software/Biotech industry during the Covid-19 pandemic. Furthermore, the industries that were most bullish in expected capital investment following the 2007-2009 financial crisis were transportation/energy and manufacturing, signaling strong demand and recovery in these sectors.

[Insert Figure 3 Here]

### 4.2. Summary Statistics

Table 1 displays descriptive statistics for the 74,429 earnings conference calls between 2006 and 2020 with non-missing financial information, which constitutes our main sample. As indicated in Panel A, a typical firm spends 1.15% on capital expenditure in a given quarter. The average *ChatGPT Investment Score* is 0.014, calculated by averaging the scores across the individual chunks from a single earnings call. Panel B compares the variables of interest across firms with high and low ChatGPT investment scores. Firms in both samples have similar company size on average, but firms with high investment scores have greater *Capital expenditure*, *Intangible investment*, *R&D expenditure*, *Total q*, and lower *Stock returns*.[11]

[11]We provide the definitions of all variables in Appendix A.

[Insert Table 1 Here]

We provide the distribution of the ChatGPT investment score in Figure 4. As described in Section 4.1, we split each conference call into text chunks of about 2,500 words to adhere to OpenAI token requirements. In subfigure (a) we see that about three-quarters of the text-chunks do not indicate any change in capital expenditure by firms, likely reflecting the fact that managers do not discuss capital expenditures in these parts of the transcripts. 11% of the text chunks show an increase, and 10.8% show a decrease in capital expending. A further 1.87% and 0.32% of the text chunks show significant increases and decreases in capital expenditure.

For each conference call, we average the text-chunk scores to obtain the ChatGPT investment score for the firm. The distribution of the firm-level score, subfigure (b), is approximately symmetric with a mode around zero. Approximately half of the firms have a non-zero investment score, suggesting that a substantial number of firm's mention plans to modify their capital investment in conference calls.

[Insert Figure 4 Here]

### 4.3. ChatGPT vs. CFO Survey Results

To the extent the ChatGPT generated score captures managerial forecasts about future corporate policies, it should be closely related to managerial beliefs on the same topics as elicited via surveys. Merging our final sample of conference call data with the CFO survey data yields a sample of 1,338 firm-quarter observations.

To visualize how the CFO survey forecasts vary across AI-predicted investment measures, we divide the matched observations into five buckets in Figure 5. As the AI-predicted investment score rises, we observe a corresponding increase in the average forecast for capital expenditure from the Duke Survey. For the subset of cases for which the AI-predicted investment measure is below -0.2, the CFO survey anticipates an average capital expenditure growth of -8.9%. In contrast, for the subset of firms with an AI-predicted investment measure above 0.2, the expected capital expenditure growth is 11.2%.

We relate the CFO Survey-based investment measure with the AI-predicted investment measure more formally in the following regression, for firm-quarter $(i, t)$,

$$CFO\ Survey\ Investment_{i,t} = \beta ChatGPT\ Investment\ Score_{i,t} + \alpha_{Ind} + \alpha_t + \epsilon_{i,t}, \tag{4}$$

where $\alpha_{Ind}$ and $\alpha_t$ are industry and time fixed effects, using the 10 industries provided in Duke CFO Survey. Table 2 reports the results.

[Insert Table 2 Here]

Table 2 shows that *CFO Survey Investment* and *ChatGPT Investment Score* are strongly positively related, statistically significance at the 1% level. Column (1) shows the R-squared from a simple regression without fixed effects is 1.4%. Column (2) shows that the correlation stays significant after including industry and time fixed effects. A one-standard-deviation increase in the *ChatGPT Investment Score* is associated with a 4.0% higher (or 0.1 standard-deviation increase in) expected capital expenditure over the next 12 months.

In summary, AI-predicted corporate policies are positively correlated with managerial beliefs, demonstrating ChatGPT's ability to extract pertinent information from large texts and the potential to complement large-scale human surveys.

## 5. ChatGPT Investment Score, Investments, and Returns

We now study ChatGPT investment score's ability to predict future investment, future returns, and the association with analyst forecasts.

### 5.1. ChatGPT Investment Score, Tobin's $q$, and Future Investments

The neoclassical theory of investment posits that Tobin's $q$ should be a sufficient statistic of firms' future investment opportunities (Hayashi, 1982). Early empirical challenges in testing the theory have been addressed by various improvements in the measurement of $q$ (e.g., Abel and Blanchard, 1986; Erickson and Whited, 2000, 2012; Philippon, 2009; Gala and Gomes, 2013; Peters and Taylor, 2017). In particular, Peters and Taylor (2017) show that the investment-$q$ relation can be substantially improved by incorporating intangible capital into the capital measurement. The total $q$ of Peters and Taylor (2017) proves to be a strong predictor of future investment activities, for both physical and intangible investment.

However, given that total $q$ still depends on the market capitalization of the firm, it might not incorporate all managerial private information about growth opportunities. Hence, the potential exists to improve the estimation of future investment opportunities through the AI-predicted investment measure, which we extract from managerial briefings. Motivated by the prediction of our model in Section 2, we examine the following regressions to study the incremental predictive power of our measure for future investments, for firm-quarter $(i, t)$,

$$\begin{aligned} \textit{Capital Expenditure}_{i,t+2} &= \beta_1 \textit{ChatGPT Investment Score}_{i,t} + \beta_2 \textit{Total } q_{i,t} \\ &\quad + \beta_3 \textit{Capital Expenditure}_{i,t} + \gamma \textit{Controls}_{i,t} + \alpha_i + \alpha_t + \epsilon_{i,t}. \end{aligned} \quad (5)$$

We include firm and time-fixed effects and cluster standard errors at the firm level. We skip quarter $t+1$ since earnings calls typically occur 30 to 60 days after the end of quarter $t$.

Table 3 shows the *ChatGPT Investment Score* positively predicts *Capital Expenditure* in the subsequent quarter, with coefficients statistically significant at 1% levels. Columns (1) to (4) demonstrate that this finding is robust to the inclusion of firm and time fixed effects and controlling for lagged capital expenditure and Total $q$ and other common determinants of firm investment. A one standard deviation increase in the *ChatGPT Investment Score* is associated with a 0.055 to 0.079 standard-deviation increase in capital expenditure in the calendar quarter following the earnings call, corresponding to 45.9% to 77.6% of the sensitivity of capital expenditure to total $q$. Therefore, the *ChatGPT Investment Score* provides substantial incremental information about firms' growth opportunities, above and beyond information in Tobin's $q$ and other common variables. In Appendix IA.1, we show the association of the *ChatGPT Investment Score* with capital expenditure holds after controling for general sentiment of the earnings call.

[Insert Table 3 Here]

Given that our prompt to ChatGPT asks about the firm's policy in the next year, we further examine whether the *ChatGPT Investment Score* has predictive power for investments at horizons longer than one quarter. In Table 4, we estimate regression (5) by replacing the dependent variable with investment in future quarters. *ChatGPT Investment Score* is positively associated with future investment for up to 9 quarters after the conference call. The coefficients are statistically significant at the 5% level or higher. The slopes for *ChatGPT Investment Score* for quarters $n = 2$ to 10 sum to 2.78%, which implies a one standard deviation increase in the *ChatGPT Investment Score* is associated with a 0.51% increase in capital expenditure in the next nine quarters, corresponding to 0.32 of a quarterly standard deviation. To alleviate concerns that our results are driven by differing firms in our sample, in Appendix Table IA.2, we show the positive association holds after keeping the sample of firms constant across different quarters.

[Insert Table 4 Here]

Peters and Taylor (2017) argue that intangible investment has become increasingly important in the economy and find Total $q$ to be a good predictor of both physical and intangible investment. Table 5 shows the *ChatGPT Investment Score* significantly and positively predicts future investment measured in different ways, including *Physical Investment*, *Intangible Investment*, *Total Investment*, and *R&D*, controlling for Total $q$. *ChatGPT Investment Score* positively predicts *Physical Investment*, *Intangible Investment*, *Total Investment*, and *R&D* in the next period, with coefficients statistically significant at 1% levels. The results are robust to the inclusion of firm and time fixed effects and controlling for Total $q$, the lagged dependent variable, and other controls. Compared to *Intangible Investment*, the predicting power of ChatGPT is larger for *Physical Investment*. Furthermore, Table IA.3 shows that the predictive power for *Total Investment* also lasts for up to 9 quarters after the earnings call. Untabulated results show similar long-term patterns for other measures of investment.

[Insert Table 5 Here]

A potential concern with the previous results is that Total $q$ may serve as a noisy proxy for true $q$ due to two primary reasons: First, intangible capital is measured with error. Second, investment theory suggests that it is the marginal $q$, rather than the average $q$, that determine investment decisions. The average $q$ can diverge from the marginal $q$, as noted in Gala (2015). To assess whether the information contained in the *ChatGPT Investment Score* is primarily captured by the true $q$, we employ the approach proposed by Erickson, Jiang, and Whited (2014), which corrects for measurement error in $q$. The results are presented in Table 6.

In column (1), we regress *Capital Expenditure* on Total $q$, whereas in column (2), we regress it on the *ChatGPT Investment Score*, acknowledging potential measurement errors in both variables. In column (3), we include both predictors in the same regression. The coefficients for both Total $q$ and the *ChatGPT Investment Score* diminish significantly when they are included together in the regression, compared to when each is included individually. Notably, the *ChatGPT Investment Score* provides a stronger explanatory power for investment compared to Total $q$, as evidenced by the higher $\rho^2$ (0.193 versus 0.106), which is the within-firm $R^2$ from a hypothetical regression of investment on true Total $q$ or true *ChatGPT Investment Score* , in column (2) relative to column (1). Columns (4) through (6) reinforce these findings when we add additional firm-level covariates to the regressions.

[Insert Table 6 Here]

Overall, the evidence indicates that our AI-based investment measure contains substantial new information for firms' growth opportunities over the short and medium term, suggesting the far-reaching impact of the expected investment measure on corporate policies and its long-term association with the environment in which companies operate. The new information appears to be largely incremental to the true Total $q$, suggesting that investors may not fully incorporate future investment policies discussed during earnings calls into stock prices in the short term.

### 5.2. Predictive Power across Information Environment

In this section, we conduct cross-sectional tests to explore the heterogeneity in the predictive power of ChatGPT. Managerial expectations and forecasts for more opaque firms and firms operating in a dynamic, changing environment could be more informative, given that these firms are subject to higher uncertainty and unexpected changes, and market prices might take longer to incorporate these plans. We consider industry competition, firm size, and product life cycle stages of firms as proxies for the environment in which a firm operates. We employ two measures for the level of competition in an industry: *HHI*, the Herfindahl-Hirschmann index, or the sum of squared market shares, in the industry defined based on textual analysis of similarities in firms' 10K product descriptions following Hoberg and Phillips (2016); *Top4Shares* is the sum of the market shares of the top four market leaders in an industry for a given quarter. The definitions of firms' product life cycle stages follow Hoberg and Maksimovic (2022), who summarize the stages of a firm's product portfolio as a four-element vector (*Life1*, *Life2*, *Life3*, *Life4*), where each component is bounded between 0 and 1 and the sum of the four components is 1. *Life1*, *Life2*, *Life3*, and *Life4* refer to the stages of product innovation, process innovation, stability and maturity, and product discontinuation, respectively. We add the interactions of the ChatGPT Investment Score with the level of competition, firm size, and stages in product life cycles to the regression to examine whether the information environment modulates the relationship between future total investment and the investment score.

Table 7 reports the results. Columns (1) to (4) show the coefficients of the interaction between *ChatGPT Investment Score* with *HHI*, *Top4Shares*, and firm size are negative and statistically significant at the 1% level, indicating that the ChatGPT investment score has a greater predictive power for future investment for small firms and firms operating in more competitive industries. Column (5) indicates the ChatGPT-based investment score is a strong predictor of future investment for firms in earlier stages of the lifecycle, i.e., the product innovation stage (*Life1*) and the process innovation stage (*Life2*), whereas it does not significantly forecast investment for firms in the mature stage (*Life3)*, and in the decline stage (*Life4*). Column (6) also controls for the interactions between *Total q* and *HHI*, *Top4Shares*, *Size*, and *Life1*-*Life4* and shows that the results remain unchanged.

[Insert Table 7 Here]

Overall, the evidence indicates that the ChatGPT-based investment scores exhibit greater power in predicting firms' future investment plans for firms in a more dynamic, changing infor-mation environment, supporting the argument that managerial forecasts are more informative for more uncertain firms.

### 5.3. ChatGPT Investment Score and Long-Term Returns

An investment factor is central in determining asset returns. The current leading factor models, the Fama-French 5-factor model (Fama and French, 2015) and the $q$-factor model (Hou, Xue, and Zhang, 2015; Hou et al., 2021), all contain an investment factor. The investment factor reflects that high-investment stocks generate lower returns than low-investment stocks. Liu, Whited, and Zhang (2009) provide theoretical foundations for the negative association of investment and expected returns. Furthermore, the expected investment growth factor in the $q$-5 factor model also indicates that it is important to estimate future investment changes. To the extent that the *ChatGPT Investment Score* captures new information regarding firms' future investment opportunities and complements the information in current investment and Tobin's $q$, our model in Section 2 predicts the *ChatGPT Investment Score* to be negatively related to future stock returns.

In Table 8, we test this hypothesis by regressing future quarterly returns on *ChatGPT Investment Score*, controlling for Total $q$ and past returns. We find the AI-predicted investment measure is negatively associated with returns over the following quarter, and the abnormal quarterly returns adjusted for the Fama-French 5-factor model and the $q$-5 factor model, with statistical significance at the 1% level. The slope of the investment score is −9.94%, −8.08%, and −7.73% for the raw return, the FF5-adjusted return, and the $q$5-adjusted return, respectively. Economically, a one-standard-deviation increase in the investment score leads to a decrease of 1.83%, 1.49%, and 1.42% in annualized return, FF5-adjusted return, and $q$5-adjusted return in the quarter subsequent to the earnings call, respectively.

[Insert Table 8 Here]

In Table IA.4 in the Appendix, we show the general sentiment of the call does not drive the return predictions. In Table IA.5, we control for contemporaneous factors from the Fama-French 5-factor model and the $q$-5 factor model and find similar results.

Table 9 further shows that the same pattern persists for up to 9 quarters for abnormal returns in the future. The negative association of *ChatGPT Investment Score* with future abnormal returns is statistically significant at the 5% or higher levels for $q5$-adjusted returns for quarters $n = 2$ to 10, and significant for FF5-adjusted returns for quarters $n = 2$ to 6 as well as $n = 9, 10$. On average, a one-standard-deviation increase in the investment score leads to a change of $-1.54\%$ in annualized $q5$-adjusted returns for each quarter $n = 2$ to 10, and $-1.10\%$ in annualized FF5-adjusted return for each quarter $n = 2$ to 6, respectively.

[Insert Table 9 Here]

The results in this section show that the ChatGPT investment score can predict long-term future returns and contains information not yet fully incorporated in standard factor models and can thus be of value to investors.

### 5.4. ChatGPT Investment Score and Short-Term Returns

In this section, we analyze whether the ChatGPT Investment Score has predictive power for short-term returns. Given that the news of higher growth opportunities and the associated lower future expected returns provide positive signals to investors, our model in Section 2 predicts a positive short-term return for earnings calls with a higher ChatGPT Investment Score. Specifically, we focus on cumulative abnormal returns in the windows [0,1], [0,3] and [0,5] days following the earnings call date. The abnormal returns are estimated from a Fama-French-Carhart 4-factor model, with betas calculated from a 100-day pre-event estimation period.

In Table 10, we regress cumulative abnormal returns on *ChatGPT Investment Score*, controlling for Total $q$ and other control variables we employed before. Managerial sentiment expressed in earning calls can also convey directional signals to investors and drive short-term stock returns. We therefore calculate the sentiment of earnings call transcripts using the Loughran and McDonald (2011) approach. In other words, we count the frequencies of negative- and positive-sentiment words based on the Loughran and McDonald (2011) dictionaries and compute the (net) sentiment as the difference of positive and negative words scaled by the total number of such words in each document. We use sentiment as a control variable in all regressions in Table 10.

[Insert Table 10 Here]

In columns (1), (3), and (5), we find the AI-predicted investment measure is positively associated with the cumulative abnormal returns for the windows [0,1], [0,3] and [0,5], with a statistical significance at the 1% level. The slope is around 3.2% for each of the regressions, suggesting that almost all of the information is incorporated into prices within one day of the conference call. Economically, a one-standard-deviation increase in the investment score leads to an increase of around 0.6% cumulative abnormal return.

In columns (2), (4), and (6), we also control for earnings surprises, which is another major factor that can drive short-term stock price responses. Since earnings surprise requires the availability of I/B/E/S analyst forecasts, the sample shrinks substantially in these specifications. We calculate earnings surprise as the change in Earnings Per Share from quarter $t-4$ to quarter $t$ divided by stock price in quarter $t$ following Livnat and Mendenhall (2006). We find the coefficient on the ChatGPT Investment Score remains virtually unchanged in all regressions.

In sum, the results indicate the ChatGPT investment score contains significant value-relevant information regarding firms' investment opportunities beyond earnings surprise and managerial sentiment, and such information is impounded into short-term stock returns by investors.

### 5.5. ChatGPT Investment Score and Analyst Forecasts

In this section, we analyze whether the ChatGPT-predicted investment score aligns with analyst forecasts. The ChatGPT measure is based on the information content of earnings call transcripts, and analysts covering a firm go over such transcripts carefully. Therefore, we expect that information in the score will be reflected in analysts' changes in capital expenditure forecasts from before to after the conference call date. In the tests of this subsection, we restrict the sample to firm-quarters for which analysts' capital expenditure forecasts exist in I/B/E/S, consisting of around half of our original sample. For a conference call that occurs in quarter $t+1$, we take the consensus in analysts' capital expenditure forecast after and within one quarter of the conference call to compute the post-call consensus forecast for capital expenditure. Similarly, we compute the pre-call consensus forecast and calculate the *Change in Analyst Forecast* as post-call minus pre-call consensus forecasts.

Table 11 shows the ChatGPT investment score is positively associated with the change in analyst forecast at the 1% significance level. The coefficient of the slope is 7.95 in the most stringent regression in column (4), with all control variables and fixed effects. Economically, a one standard deviation increase in the ChatGPT investment score is associated with a 1.46% increase in analyst capital expenditure forecast.

These tests validate that the ChatGPT investment score captures important information regarding firms' future capital expenditure plans, which is reflected in analysts' forecast revisions.

[Insert Table 11 Here]

### 5.6. Factors that Drive Expected Investment

In this section, we further analyze different factors that drive firms' expected investment as reflected in the ChatGPT investment score. We start with the simple explanations provided by ChatGPT in response to our prompt in Section 4.1. We gather all explanations and ask ChatGPT to summarize them into different topics. After reviewing the summarized topics, we consolidate them into three broad categories: (1) *Operational Efficiency and Cost Control*, including efficiency improvement, cost reduction, resource reallocation, infrastructure resilience enhancement, etc.; (2) *Strategic Planning and Business Focus*, including strategic acquisitions and partnerships, revenue growth support, expansions of operations, strategic restructuring, focus shift, etc.; and (3) *Regulatory Compliance and Market Conditions*, including regulatory compliance, safety investments, economic and market conditions, pandemic adjustments, etc.

[Insert Figure 6 Here]

Figure 6 illustrates the temporal dynamics and implications of different investment explanation categories. Panel A reveals distinct patterns in how firms justify their investment decisions across economic cycles. During economic downturns (especially during 2008-2009 and 2020), firms cite “Regulatory Compliance and Market Conditions” and “Operational Efficiency and Cost Control” as primary drivers of investment decisions. This pattern reflects firms’ strategic responses to challenging economic environments. In contrast, “Strategic Planning and Business Focus” explanations dominate during periods of economic stability.

Panel B examines the relationship between these explanation categories and firms’ investment intentions. When firms cite “Regulatory Compliance and Market Conditions” or “Operational Efficiency and Cost Control,” they typically signal investment reductions, aligning with their defensive posture during economic stress. Conversely, firms that emphasize “Strategic Planning and Business Focus” overwhelmingly indicate plans to increase investments, consistent with a growth-oriented mindset under stable economic conditions.

Table 12 presents the regression results examining the drivers of ChatGPT’s investment assessments. To evaluate the predictive power of different explanation categories, we interact the ChatGPT Investment Score with indicator variables for three distinct categories of investment rationale. The results indicate that investment scores in all three categories predict long-term capital expenditure up to nine quarters in the future. Investments related to “Regulatory Compliance and Market Conditions” have a shorter predictive horizon of seven quarters, consistent with the idea that responses to adverse market conditions may be more transitory in nature. Untabulated results also show that investment scores in all three categories also predict both immediate market reactions (measured by cumulative abnormal returns around conference calls) and long-term stock returns.

[Insert Table 12 Here]

## 6. Robustness Tests and Additional Analyses

This section conducts robustness tests of the previous results, and shows the effectiveness of ChatGPT for understanding other corporate policies.

### 6.1. Out-of-sample and Masked-identity Tests

In this section, we address the concerns that ChatGPT may use public information other than the content in a particular earnings call, as it is trained with large-scale public datasets, which could result in look-ahead bias. We conduct two separate tests to alleviate this concern: (i) we rerun our analysis on a subsample of conference calls that happened after the ChatGPT training period; (ii) we mask the identities of firms, managers, and products in conference call transcripts, and re-run the prompts with ChatGPT to generate investment scores.

First, we regenerate the ChatGPT investment score and reconduct our main analysis for the subsample of earnings calls that occurred after the training period of the ChatGPT model. Since the ChatGPT 3.5 model's training data includes information up until September 2021, the model cannot be aware of events or information after that date. Therefore, we conduct the analysis for the 2021Q4 to 2022Q4 period, which would be void of any look-ahead bias.

Table IA.6 shows that our main results for future investment are robust in this restricted sample. In Table IA.6, the ChatGPT Investment Score positively predicts capital expenditure in the next period, with coefficients statistically significant at 1% levels. Columns (1) to (4) demonstrate that this finding is robust to the inclusion of firm and time fixed effects and controlling for Total $q$ and other predictors of investment. A one standard deviation increase in *ChatGPT Investment Score* is associated with a 0.084 to 0.133 standard deviation increase in capital expenditure in the calendar quarter following the earnings call.

Next, we return to the same sample period as for our main results – 2006 to 2020. However, we mask the identity of words that could reveal the identity of a firm. We anonymize the text by removing dates, personal names, organization names, and product names. To achieve this, we process the conference call text using regular expressions to identify years (from 1900 to 2099) and month names (including their abbreviations). Furthermore, we use a general English language parsing model developed by spaCy[12] to tag nouns in conference calls that correspond to firm, people, and product names. Once identified, each piece of identifying information is replaced with "###". To economize on the costs of running the task, we restrict the analysis to a random 10% subsample of our original final sample.

[12]The model package is available from the (en_core_web_sm) model at https://spacy.io/models/en.

Table IA.7 shows that our main results hold after removing identifying information from the earnings call transcripts. The slope of the ChatGPT investment score after including all control variables and fixed effects is 0.224, which is slightly lower than the 0.477 reported in our main results in Table 3. However, this reduction is primarily due to the different sample, as the coefficient with our original measure is 0.323 in the subsample, according to unreported results.

### 6.2. Additional Corporate Controls

In previous sections, we control for *Total q*, calculated using the market value from the quarter before the earnings call date, and common firm covariates that influence future capital expenditure, including past *Capital Expenditure*, *Total Cash Flow*, *Leverage*, *Size*. However, the market value immediately following an earnings call could capture information from the call, including forecasts for capital expenditure (as shown in Section 5.4). Additionally, firms may have longer-term investment plans and thus capital expenditures from earlier periods can be correlated with future investment. Furthermore, other firm-level characteristics, such as profitability, sales growth, and financial constraints, might also affect firms' future capital expenditure plans.

To address these concerns, we add more control variables to our regression analysis for predicting future capital expenditure plans. These include *Total q* calculated with the updated market values at 0, 1, or 5 days after the earnings call, *Profitability*, *Sales Growth*, *Z-score*, and eight lags of capital expenditure, as detailed in Table IA.8.

Table IA.8 demonstrates that our main results remain consistent even after including all of these additional covariates. The slope of the *ChatGPT Investment Score* varies from 0.289 to 0.291, which is slightly lower than the coefficient of 0.477 in our primary analysis in Table 3. The results suggest that the *ChatGPT Investment Score* captures information about firms' future investment plans that is not yet reflected in their stock prices and cannot be fully explained by other observable firm covariates.

### 6.3. Other Large Language Models

Although ChatGPT represents a significant advancement in natural language processing, earlier models are also quite powerful. We test in this section whether other large language models (LLMs) can also be used to interpret corporate policies. One such model – RoBERTa, or Robustly optimized BERT approach, is a transformer-based neural network model for natural language processing. It was introduced by Facebook AI Research (Liu et al., 2019) as an extension and improvement upon Google's BERT model. RoBERTa was the leading LLM until the release of the larger and more capable GPT models.

We start with the pre-trained RoBERTa model and fine-tune the model by training it on question-answering tasks with the BoolQ open-source dataset of questions and answers, available through HuggingFace.[13] Since the token limit for RoBERTa is 512, we split conference call transcripts into 300-word text chunks. Finally, we query RoBERTa with the question "Does the firm plan to increase its capital spending over the next year?" for each text chunk, and average the scores for each conference call. The average RoBERTa investment score is 0.220 with a standard deviation of 0.064 (compared to 0.014 and 0.184 respectively for ChatGPT Investment Score).

Column (2) of Table IA.9 shows that the RoBERTa model also does well in forecasting investment. The RoBERTa investment score is positively associated with the firm's capital expenditure with a coefficient of 1.321, significant at the 1% level. Column (3) conducts a horse race regression with both the ChatGPT score and the Robert score included. The coefficient for the RoBERT score reduces to 0.98, an almost 26% drop, whereas the coefficient on the ChatGPT score only declines by 7%. The statistical significance for the coefficient of the ChatGPT score (12.99) is also much higher than that of the RoBERTa score (6.23). Further, for a one-standard-deviation increase in the RoBERTa score, the capital expenditure increases by a 0.039 of a standard-deviation, compared with 0.051 of a standard-deviation for the ChatGPT score. Therefore, both the economic and statistical significance for the RoBERTa model are less pronounced than those for the ChatGPT score, implying that employing more advanced LLM models yields greater informational

[13] Each example in the dataset consists of a passage, a question, and a Boolean answer (either "true" or "false"); Link: https://huggingface.co/datasets/boolq.

content.

### 6.4. Alternative ChatGPT Scores

In this section, we consider an alternative definition of the ChatGPT investment score, *ChatGPT Investment Alt. Score*, in which we take the largest value of the ChatGPT answers among all chunks of an earnings call. Specifically, we take the text-chunk with the greatest absolute value of ChatGPT-assigned investment score, and assign the corresponding signed score to the conference call. If there are two text-chunks with extreme investment scores with equal absolute value but opposite signs, we assign 0 to the conference call. This measure can be justified on the ground that the most salient information conveyed by the manager in the entire earnings call should be used to define the score. Table IA.10 shows that our main results for future investment and returns are robust to this measure. In Table IA.10, *ChatGPT Investment Alt. Score* positively predicts *Capital Expenditure* in the next period, with coefficients statistically significant at 1% levels. Columns (1) to (4) demonstrate that this finding is robust to the inclusion of firm and time fixed effects and controlling for lagged capital expenditure and Total $q$. A one standard deviation increase in *ChatGPT Investment Alt. Score* is associated with 0.034 to 0.046 standard deviation increase in capital expenditure in the calendar quarter following the earnings call.

As the inference of ChatGPT involves some degree of randomness, we also consider whether different attempts of ChatGPT yield different scores. For this purpose, we run the ChatGPT model on a 10% subsample of our sample two times and calculate the corresponding ChatGPT investment scores. The correlation of the two scores is extremely high at 98.68%, and they generate quantitatively similar inferences in untabulated results.

The results in this section provide further evidence that ChatGPT can predict firms' future capital expenditure, and the precise way of how we construct our firm-level measure from chunk-level responses does not make any material difference.

### 6.5. ChatGPT and Other Corporate Policies

So far, we have focused on firms' investment policy. The methodology we develop, however, can be equally applied to extract firms' expectations about other corporate policies. We consider two important discretionary policies: dividend payment and hiring. We follow the method described in Section 4.1, but replace "capital spending" with "dividend payment" and "employment," respectively, to construct a *ChatGPT dividend score* and a *ChatGPT employment Score*. Specifically, we input the following prompt into the model.

> The following text is an excerpt from a company's earnings call transcripts. You are a finance expert. Based on this text only, please answer the following questions. 1. How does the firm plan to change its dividend payment over the next year? 2. How does the firm plan to change its number of employees over the next year? There are five choices: Increase substantially, increase, no change, decrease, and decrease substantially. Please select one of the above five choices for each question and provide a one-sentence explanation of your choice for each question. The format for the answer to each question should be "choice - explanation." If no relevant information is provided related to the question, answer "no information is provided. Please answer each question independently."
>
> *[Part of an earnings call transcript.]*

The ChatGPT model provides a combination of choice-explanation for the two questions separately. For each question, we assign a score of -1, -0.5, 0, 0.5, and 1 for each of the given choices (decrease substantially, decrease, no change, increase, and increase substantially), respectively. If ChatGPT generates an answer "no information is provided," we assign a value of zero to the score. We then take the average of the scores across multiple chunks of one earnings call to obtain a firm-quarter-level measure of *ChatGPT Dividend Score* and *ChatGPT Employment Score*.

Table 13 validates that *ChatGPT Dividend Score* and *ChatGPT Employment Score* are significantly and positively associated with the answers to the Duke CFO Surveys. Columns (1) and (3) show that the R-squared from a simple dividend or employee regression is 2.3% and 0.7% without fixed effects. Column (2) and Column (4) show that the correlation stays significant at the 1% level after including industry and time fixed effects. A one standard deviation increase in the *ChatGPT Dividend Score* is associated with a 0.11 standard deviation increase in the *CFO Survey Dividend* response. A one-standard-deviation increase in *ChatGPT Employment Score* is associated with a 0.07 standard-deviation increase in the *CFO Survey Employment* answer.

Combined with our previous findings, Table 13 adds supportive evidence that ChatGPT can extract valuable information regarding corporate policies from earnings conference calls and has the potential to complement traditional surveys of corporate executives.

[Insert Table 13 Here]

## 7. Concluding Remarks

In this paper, we use the cutting-edge large language model, ChatGPT, to extract managerial expectations of corporate policies from corporate disclosure. We construct a ChatGPT investment score that measures the extent to which managers expect to increase or decrease capital expenditures in the future. The ChatGPT investment score is supported by interpretable textual content and is strongly correlated with survey responses from CFOs. We build a stylized model to guide our empirical analyses. Consistent with the model prediction, the investment score bears a strong, positive correlation with future investment both in the short term and long term, even after controlling for Tobin's $q$ and other determinants of investment, indicating that managers convey new information about firms' future investment opportunities in conference calls that ChatGPT helps to extract. The new information conveyed by managers has a larger predictive ability when firms operate in an environment that is more opage, dynamic and subject to change. Furthermore, firms with high investment scores experience significantly negative future long-term abnormal returns, and significantly positive short-term returns around the conference call dates, consistent with our model's predictions.

We conducted several robustness checks to validate the results. Additionally, we extended our analysis to other corporate policies, namely dividend payment and hiring, and find that ChatGPT can effectively extract firms' expectations regarding these policies as well. Our study provides a first look at the potential of ChatGPT to extract managerial expectations and corporate policies. We believe that our findings have important implications for companies, investors, policymakers, and researchers.

Our findings have several implications. First, they suggest that ChatGPT can be used to extract valuable information about corporate policies that is not otherwise immediately available to investors. Second, they demonstrate that ChatGPT can be used to improve the predictions of future investment and returns. Third, our approach can be used to expand and complement traditional surveys of executives. Fourth, we provide a new application of AI that produce interpretable outputs for humans.

## References

Abel, A. B., and O. J. Blanchard. 1986. The present value of profits and cyclical movements in investment. *Econometrica* 54:249–73.

Abel, A. B., and J. C. Eberly. 1994. A unified model of investment under uncertainty. *American Economic Review* 84:1369–84.

Acikalin, U., T. Caskurlu, G. Hoberg, and G. M. Phillips. 2022. Intellectual Property Protection Lost and Competition: An Examination Using Large Language Models. doi:10.2139/ssrn.4023622.

Andrei, D., W. Mann, and N. Moyen. 2019. Why did the q theory of investment start working? *Journal of Financial Economics* 133:251–72.

Armour, J., R. Parnham, and M. Sako. 2022. Augmented lawyering. *University of Illinois Law Review* 71.

Bakke, T.-E., and T. M. Whited. 2010. Which firms follow the market? an analysis of corporate investment decisions. *Review of Financial Studies* 23:1941–80.

Bolton, P., H. Chen, and N. Wang. 2013. Market timing, investment, and risk management. *Journal of Financial Economics* 109:40–62.

Bond, P., A. Edmans, and I. Goldstein. 2012. The real effects of financial markets. *Annual Review of Financial Economics* 4:339–60.

Brogaard, J., M. C. Ringgenberg, and D. Roesch. 2023. Does floor trading matter? *Journal of Finance* forthcoming.

Bybee, L., B. T. Kelly, A. Manela, and D. Xiu. 2023. Business news and business cycles. *Journal of Finance* forthcoming.

Caballero, R. J. 1999. Aggregate investment. *Handbook of macroeconomics* 1:813–62.

Candia, B., M. Weber, Y. Gorodnichenko, and O. Coibion. 2023. Perceived and expected rates of inflation of us firms. In *AEA Papers and Proceedings*, vol. 113, 52–5. American Economic Association.

Cao, S., W. Jiang, J. L. Wang, and B. Yang. 2022. From man vs. machine to man + machine: The art and AI of stock analyses. *Journal of Financial Economics* forthcoming.

Cao, S., W. Jiang, B. Yang, and A. L. Zhang. 2023. How to talk when a machine is listening?: Corporate disclosure in the age of AI. *Review of Financial Studies* forthcoming.

Chen, Q., I. Goldstein, and W. Jiang. 2007. Price informativeness and investment sensitivity to stock price. *The Review of Financial Studies* 20:619–50.

Chen, Y., B. T. Kelly, and D. Xiu. 2023. Expected returns and large language models. Working Paper.

Coibion, O., Y. Gorodnichenko, and S. Kumar. 2018. How do firms form their expectations? new survey evidence. *American Economic Review* 108:2671–713.

Coibion, O., Y. Gorodnichenko, and T. Ropele. 2020. Inflation expectations and firm decisions: New causal evidence. *The Quarterly Journal of Economics* 135:165–219.

Coibion, O., Y. Gorodnichenko, and M. Weber. 2022. Monetary policy communications and their effects on household inflation expectations. *Journal of Political Economy* 130:1537–84.

Cong, L. W., T. Liang, X. Zhang, and W. Zhu. 2019. Textual factors: A scalable, interpretable, and data-driven approach to analyzing unstructured information. Working Paper.

Corrado, C. A., and C. R. Hulten. 2010. How do you measure a "technological revolution"? *American Economic Review* 100:99–104.

———. 2014. Innovation accounting. In *Measuring Economic Sustainability and Progress*, 595–628. University of Chicago Press.

D'Acunto, F., D. Hoang, and M. Weber. 2022. Managing households' expectations with unconventional policies. *The Review of Financial Studies* 35:1597–642.

Eisfeldt, A. L., and D. Papanikolaou. 2013. Organization capital and the cross-section of expected returns. *The Journal of Finance* 68:1365–406.

———. 2014. The value and ownership of intangible capital. *American Economic Review* 104:189–94.

Erickson, T., C. H. Jiang, and T. M. Whited. 2014. Minimum distance estimation of the errors-in-variables model using linear cumulant equations. *Journal of Econometrics* 183:211–21.

Erickson, T., and T. Whited. 2012. Treating measurement error in tobin's q. *Review of Financial Studies* 25:1286–329.

Erickson, T., and T. M. Whited. 2000. Measurement error and the relationship between investment and q. *Journal of Political Economy* 108:1027–57.

Ewens, M., R. H. Peters, and S. Wang. 2019. Measuring intangible capital with market prices. Working Paper, National Bureau of Economic Research.

Falato, A., D. Kadyrzhanova, J. Sim, and R. Steri. 2022. Rising intangible capital, shrinking debt capacity, and the us corporate savings glut. *The Journal of Finance* 77:2799–852.

Fama, E. F., and K. R. French. 2015. A five-factor asset pricing model. *Journal of Financial Economics* 116:1–22.

Florackis, C., C. Louca, R. Michaely, and M. Weber. 2023. Cybersecurity risk. *The Review of Financial Studies* 36:351–407.

Gala, V. D. 2015. Measuring marginal q. *Allied Social Science Associations Program* .

Gala, V. D., and J. F. Gomes. 2013. Beyond q: investment without asset prices. Working Paper, London Business School and the Wharton School of the University of Pennsylvania.

Goldstein, I. 2023. Information in financial markets and its real effects. *Review of Finance* 27:1–32.

Graham, J. R., J. Grennan, C. R. Harvey, and S. Rajgopal. 2022. Corporate culture: Evidence from the field. *Journal of Financial Economics* 146:552–93.

Graham, J. R., and C. R. Harvey. 2001. The theory and practice of corporate finance: Evidence from the field. *Journal of Financial Economics* 60:187–243.

Graham, J. R., C. R. Harvey, and M. Puri. 2013. Managerial attitudes and corporate actions. *Journal of financial economics* 109:103–21.

Hanley, K. W., and G. Hoberg. 2010. The information content of ipo prospectuses. *The Review of Financial Studies* 23:2821–64.

Hansen, A. L., and S. Kazinnik. 2023. Can ChatGPT decipher Fedspeak? Working Paper.

Hassan, T. A., S. Hollander, L. Van Lent, and A. Tahoun. 2019. Firm-level political risk: Measurement and effects. *The Quarterly Journal of Economics* 134:2135–202.

Hassett, K. A., and R. G. Hubbard. 1997. Tax policy and investment. *Fiscal policy: Lessons from economic research* 339–85.

Hayashi, F. 1982. Tobin's marginal q and average q: A neoclassical interpretation. *Econometrica* 50:213–24.

Hoberg, G., and V. Maksimovic. 2022. Product life cycles in corporate finance. *The Review of Financial Studies* 35:4249–99.

Hoberg, G., and G. Phillips. 2010. Product market synergies and competition in mergers and acquisitions: A text-based analysis. *The Review of Financial Studies* 23:3773–811.

———. 2016. Text-based network industries and endogenous product differentiation. *Journal of Political Economy* 124:1423–65.

Hou, K., H. Mo, C. Xue, and L. Zhang. 2021. An augmented q-factor model with expected growth. *Review of Finance* 25:1–41.

Hou, K., C. Xue, and L. Zhang. 2015. Digesting anomalies: An investment approach. *Review of Financial Studies* 28:650–705.

Jha, M., H. Liu, and A. Manela. 2021. Natural disaster effects on popular sentiment toward finance. *Journal of Financial and Quantitative Analysis* 56:2584–604.

———. 2022. Does finance benefit society? A language embedding approach. Working Paper.

Jiang, F., J. Lee, X. Martin, and G. Zhou. 2019. Manager sentiment and stock returns. *Journal of Financial Economics* 132:126–49.

Kim, A. G., M. Muhn, and V. V. Nikolaev. 2023. Bloated disclosures: Can ChatGPT help investors process information? Working Paper.

Li, K., F. Mai, R. Shen, and X. Yan. 2021. Measuring corporate culture using machine learning. *The Review of Financial Studies* 34:3265–315.

Liu, L. X., T. M. Whited, and L. Zhang. 2009. Investment-based expected stock returns. *Journal of Political Economy* 117:1105–39.

Liu, Y., M. Ott, N. Goyal, J. Du, M. Joshi, D. Chen, O. Levy, M. Lewis, L. Zettlemoyer, and V. Stoyanov. 2019. Roberta: A robustly optimized bert pretraining approach .

Livnat, J., and R. R. Mendenhall. 2006. Comparing the post–earnings announcement drift for surprises calculated from analyst and time series forecasts. *Journal of accounting research* 44:177–205.

Lopez-Lira, A., and Y. Tang. 2023. Can ChatGPT forecast stock price movements? return predictability and large language models. Working Paper.

Loughran, T., and B. McDonald. 2011. When is a liability not a liability? textual analysis, dictionaries, and 10-ks. *Journal of Finance* 66:35–65.

———. 2016. Textual analysis in accounting and finance: A survey. *Journal of Accounting Research* 54:1187–230.

McConnell, J. J., and C. J. Muscarella. 1985. Corporate capital expenditure decisions and the market value of the firm. *Journal of Financial Economics* 14:399–422.

Peters, R. H., and L. A. Taylor. 2017. Intangible capital and the investment-q relation. *Journal of Financial Economics* 123:251–72.

Philippon, T. 2009. The bond market's q. *The Quarterly Journal of Economics* 124:1011–56.

Pickert, R. 2023. Survey response rates are down since Covid: That's worrying for US economic data. *Bloomberg* February 15, 2023.

Tetlock, P. C. 2007. Giving content to investor sentiment: The role of media in the stock market. *The Journal of Finance* 62:1139–68.

Titman, S., K. J. Wei, and F. Xie. 2004. Capital investments and stock returns. *Journal of Financial and Quantitative Analysis* 39:677–700.

Tobin, J. 1969. A general equilibrium approach to monetary theory. *Journal of Money, Credit and Banking* 1:15–29.

Weber, M., F. D'Acunto, Y. Gorodnichenko, and O. Coibion. 2022. The subjective inflation expectations of households and firms: Measurement, determinants, and implications. *Journal of Economic Perspectives* 36:157–84.

Yang, S. 2023. Predictive patentomics: Forecasting innovation success and valuation with ChatGPT. Working Paper.

**Figure 1.** Word Clouds for Texts with High and Low ChatGPT Investment Score

This figure represents important bigrams associated with the ChatGPT investment score. We document the 25 most frequent bigrams associated with conference call texts with high and low ChatGPT investment scores. We lemmatize each word to account for differing grammatical noun and verb forms. We also exclude stop words and bigrams that contain time-related words, such as "year," "quarter," etc. More frequent bigrams are shown with larger text fonts.

**(a)** Bigrams associated with low ChatGPT investment scores.

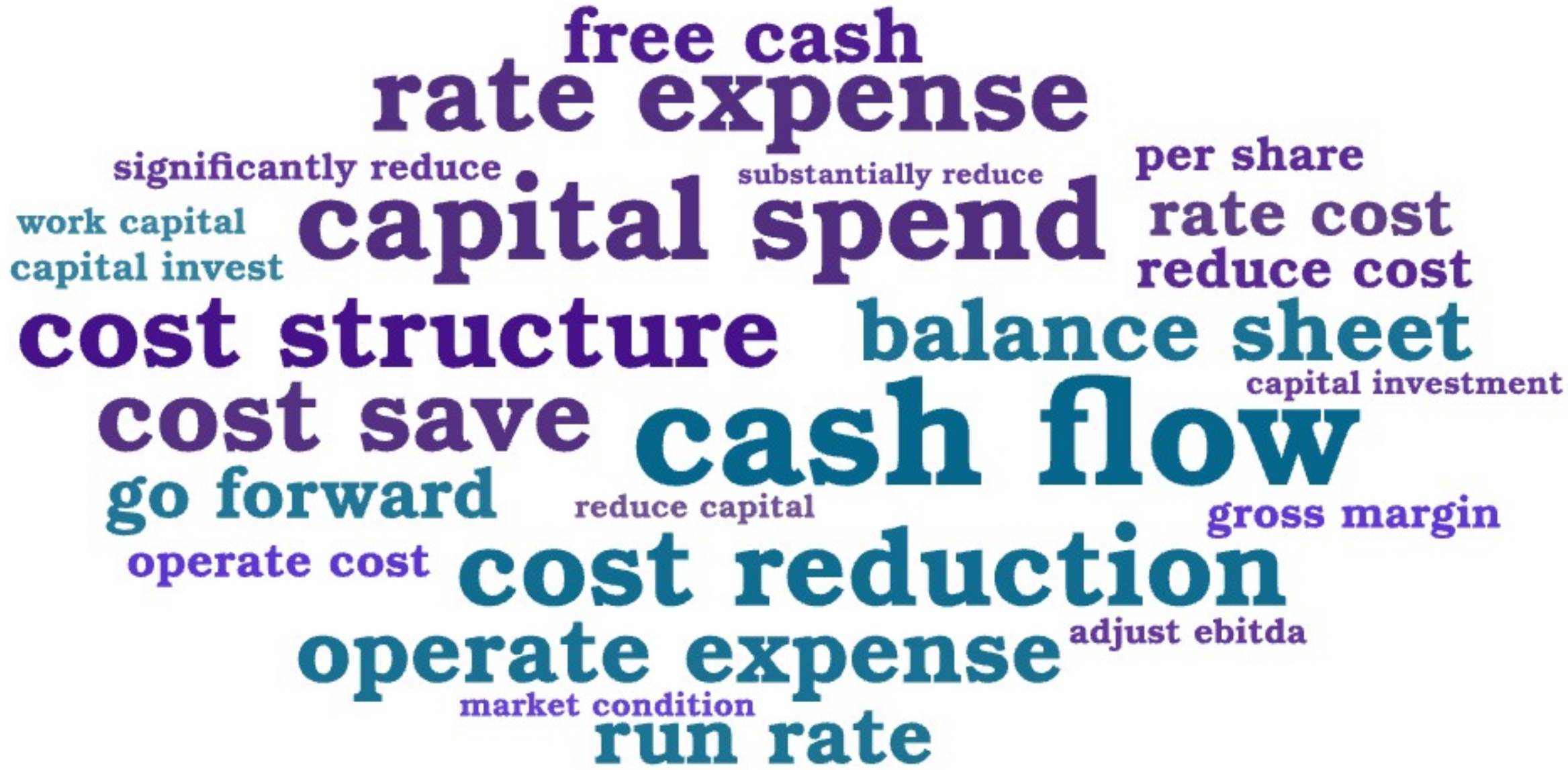

**(b)** Bigrams associated with high ChatGPT investment scores.

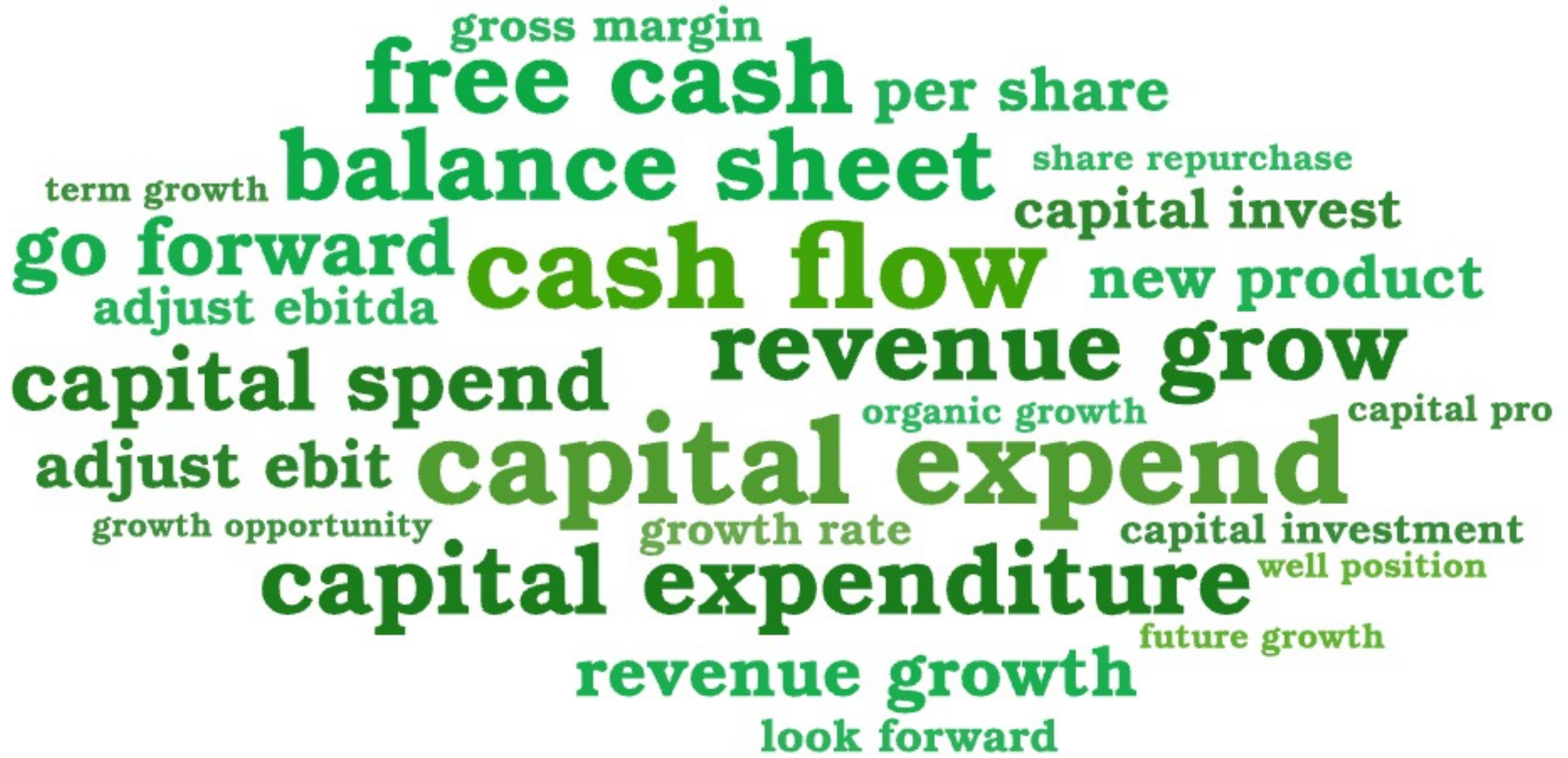

**Figure 2.** ChatGPT Investment Score vs. Realized Investment

This figure shows the time series of the average quarterly ChatGPT investment score and average future four-quarter change in capital expenditure. ChatGPT investment score is calculated based on conference call texts of the firm (described in Section 4.1). We calculate the change in capital expenditure as the difference between the average capital expenditure for the four quarters following the current quarter (t+1 to t+4) and the average capital expenditure for the four quarters prior to the current quarter (t-4 to t-1).

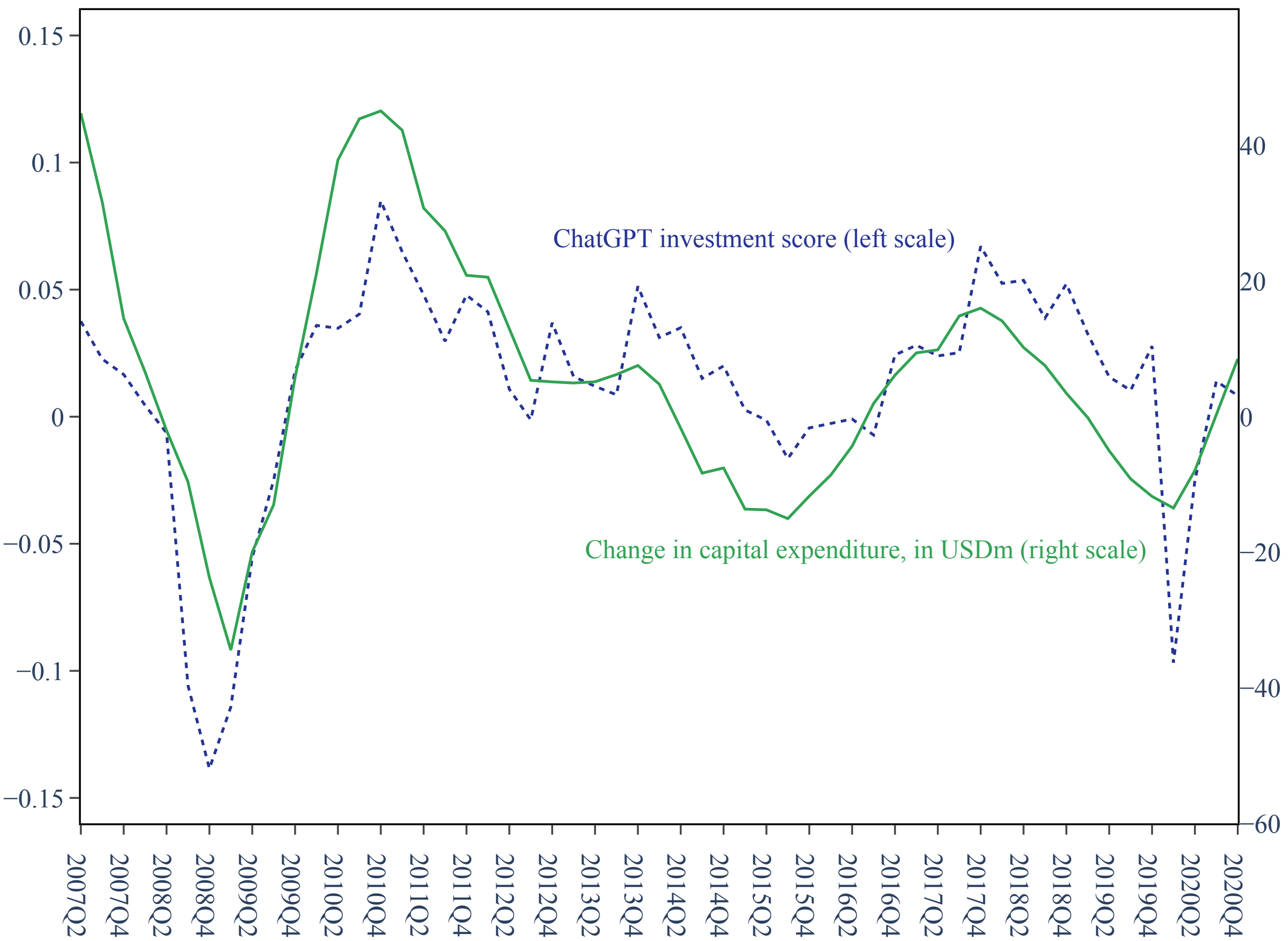

**Figure 3.** ChatGPT Investment Score across Industries

This figure represents average yearly ChatGPT investment score across industries. ChatGPT investment score is calculated based on conference call texts of the firm (described in Section 4.1). The firms are aggregated into ten industries, following the Duke CFO survey (Graham and Harvey, 2001).

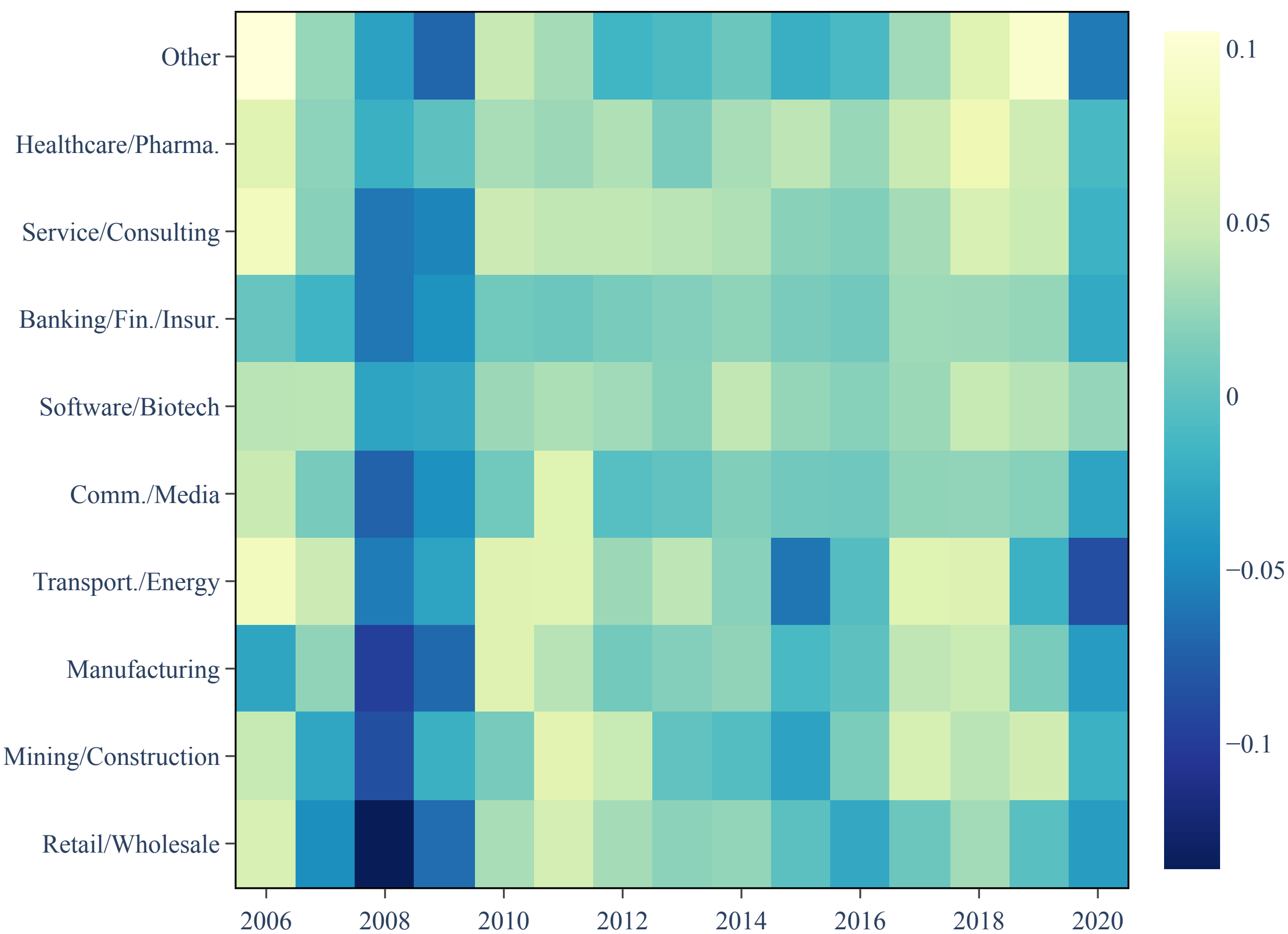

**Figure 4.** Distribution of ChatGPT Investment Score

This figure presents the distributions of ChatGPT investment score across text-chunks and conference calls. Each conference call is divided into text-chunks of length around 2,500 words (usually three to four text-chunks per conference call), to accommodate the ChatGPT's token limit. We average the score across text-chunks to obtain the ChatGPT investment score for the conference call.

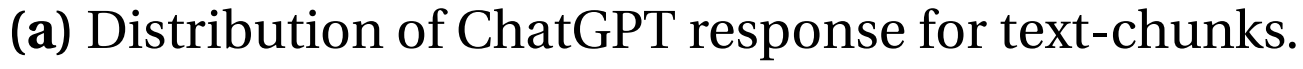

**(a)** Distribution of ChatGPT response for text-chunks.

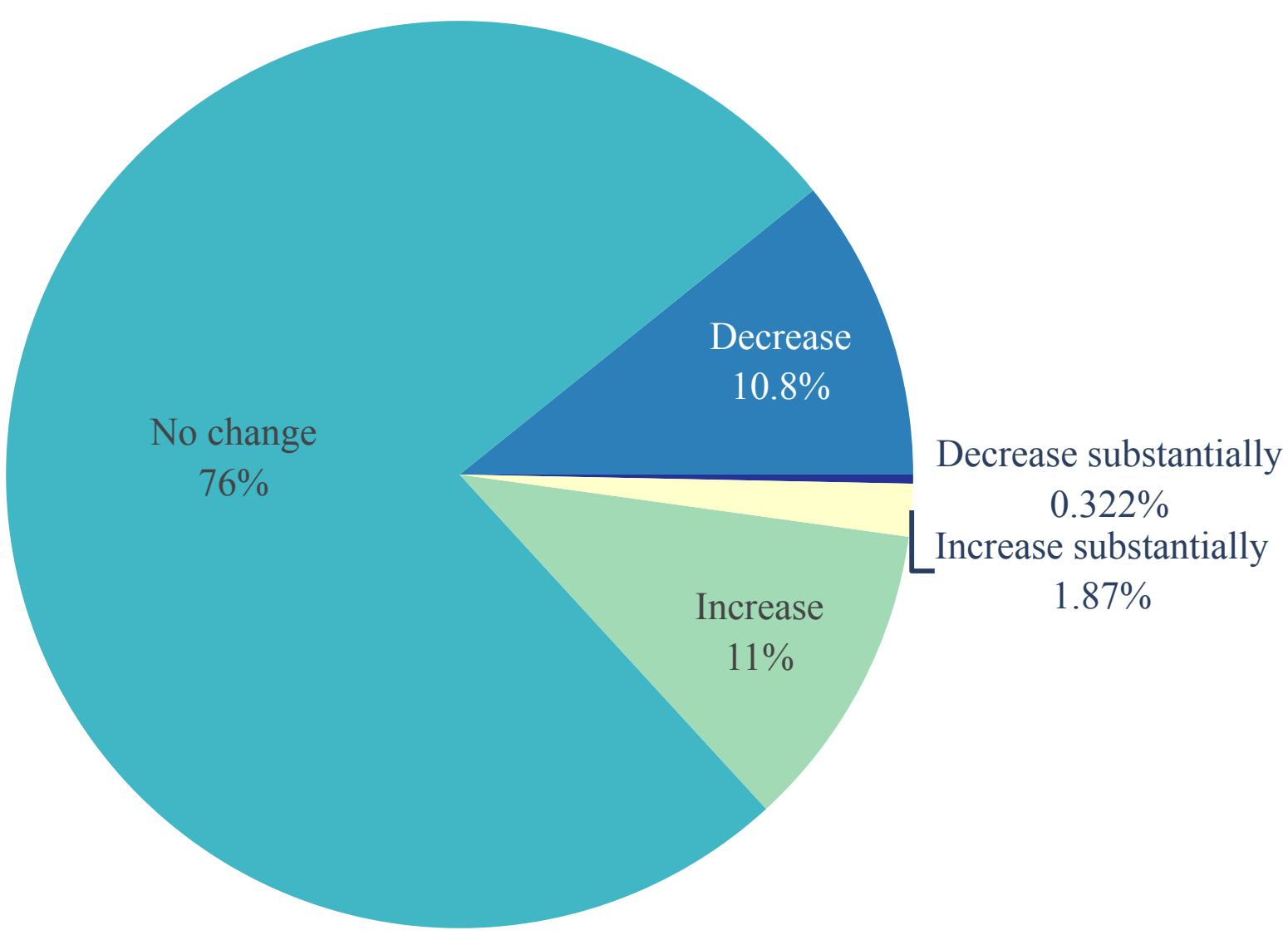

**(b)** Distribution of ChatGPT investment score for conference calls.

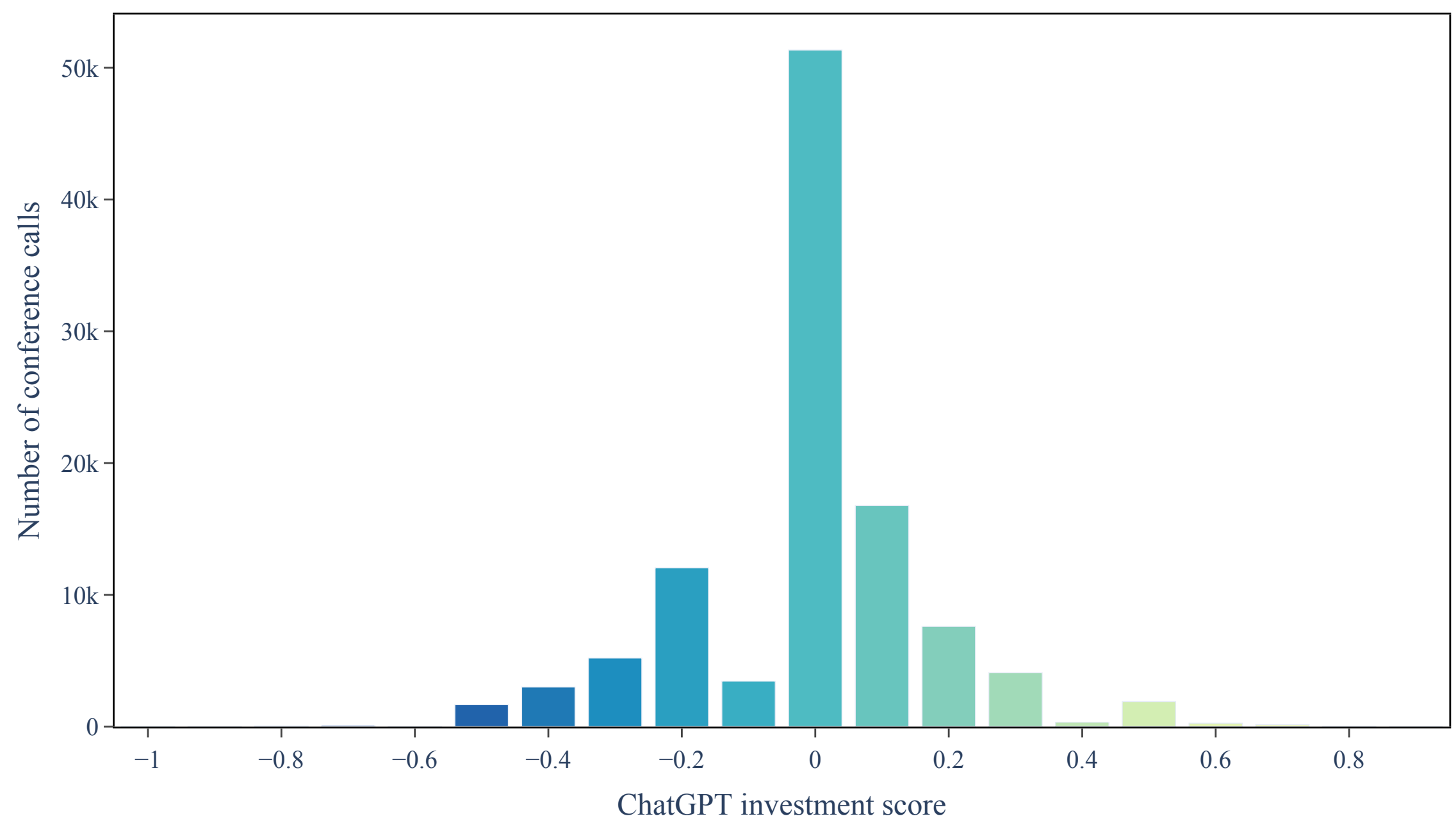

**Figure 5.** Average Capex Growth Forecast across ChatGPT Scores

This figure presents the average forecast of capital expenditure growth as reported in Duke CFO survey (Graham and Harvey, 2001) for different levels of ChatGPT investment scores. We divide our matched sample into five buckets in two steps: (i) we keep observations with ChatGPT score = 0 in the middle bucket, which consists of 51% of the sample; (ii) we divide all observations with positive and negative ChatGPT score observations into two buckets of similar numbers of observations, respectively. There are 146, 191, 681, 180, and 140 observations in the five buckets from left to right, respectively.

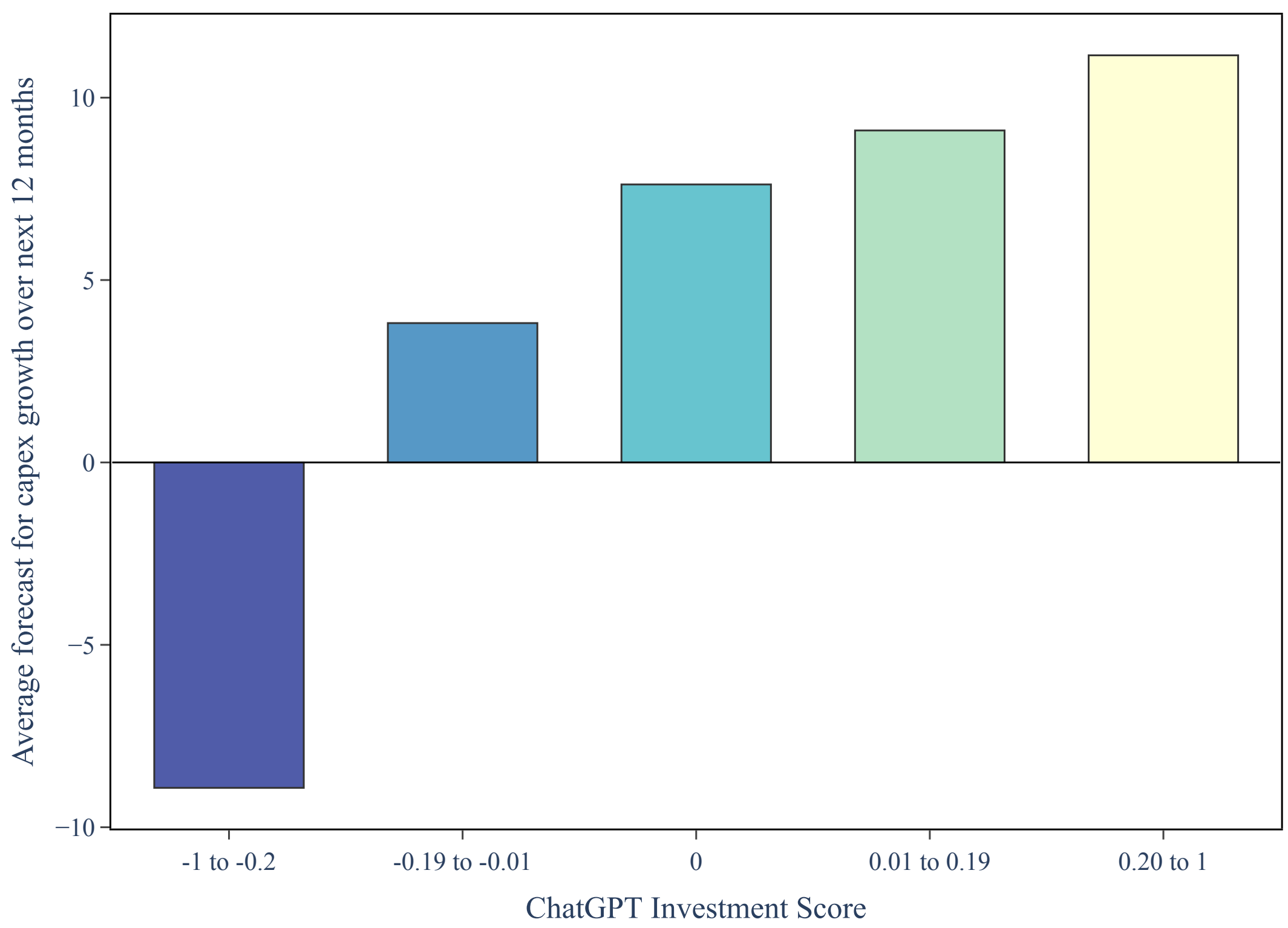

**Figure 6.** Factors that Drive Expected Investment

This figure presents factors that drive ChatGPT's assessments of expected investments based on the one-sentence explanations provided by ChatGPT. Panel (a) shows the temporal evoluation of the shares of explanations that fall in three broad categories: (i) Operational Efficiency and Cost Control, (ii) Strategic Planning and Business Focus, and (iii) Regulatory Compliance and Market Conditions. Each colored area corresponds to the explanations associated with one category. Panel (b) displays the share of explanations in each category that is associated with a positive ChatGPT Investment Score, i.e., associated with an expected increase in investment. Each line plots the fraction of explanations for one category.

**(a)** Distribution of ChatGPT explanations across three broad categories.

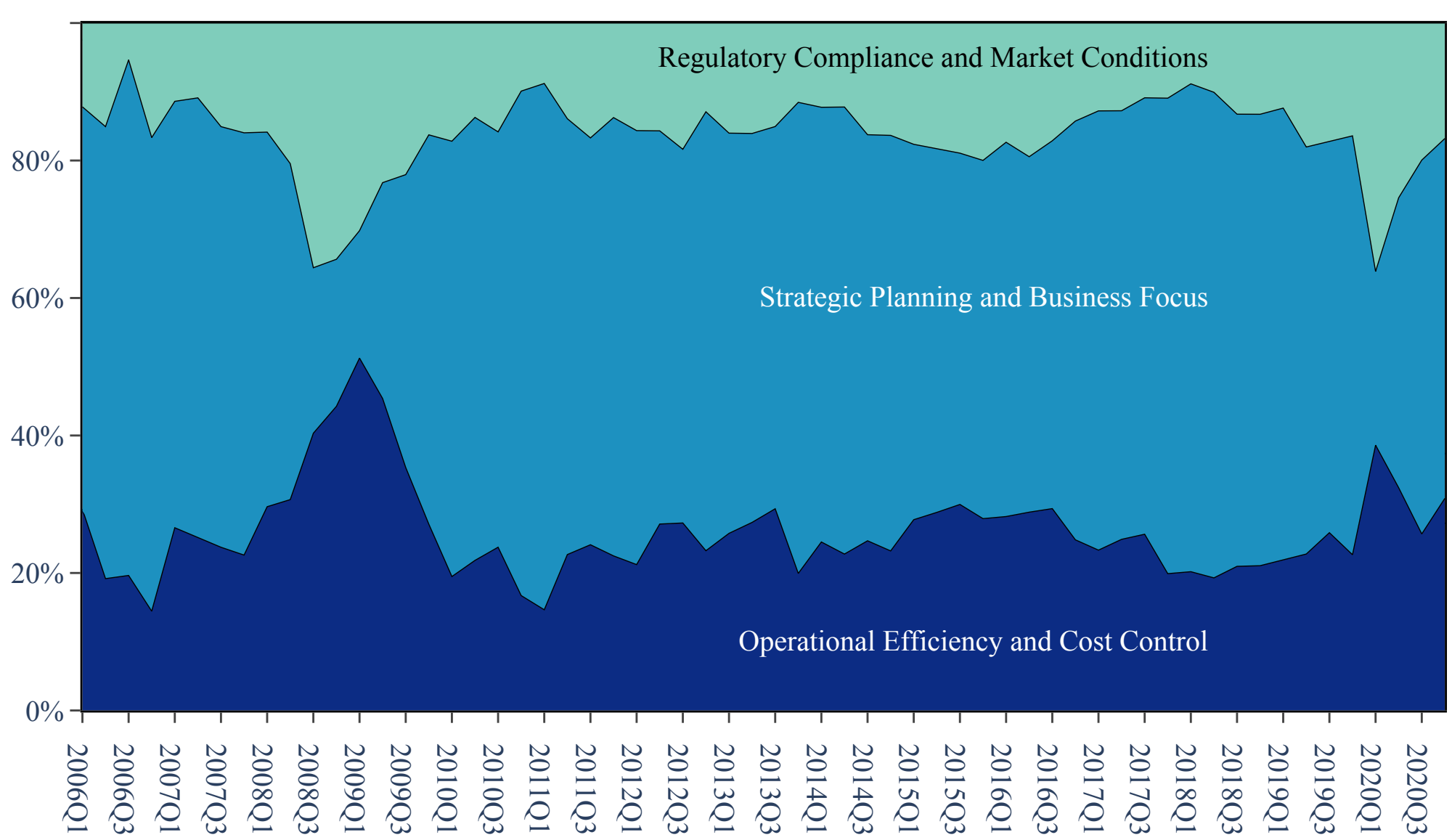

**(b)** Share of positive investment assessments in each category.

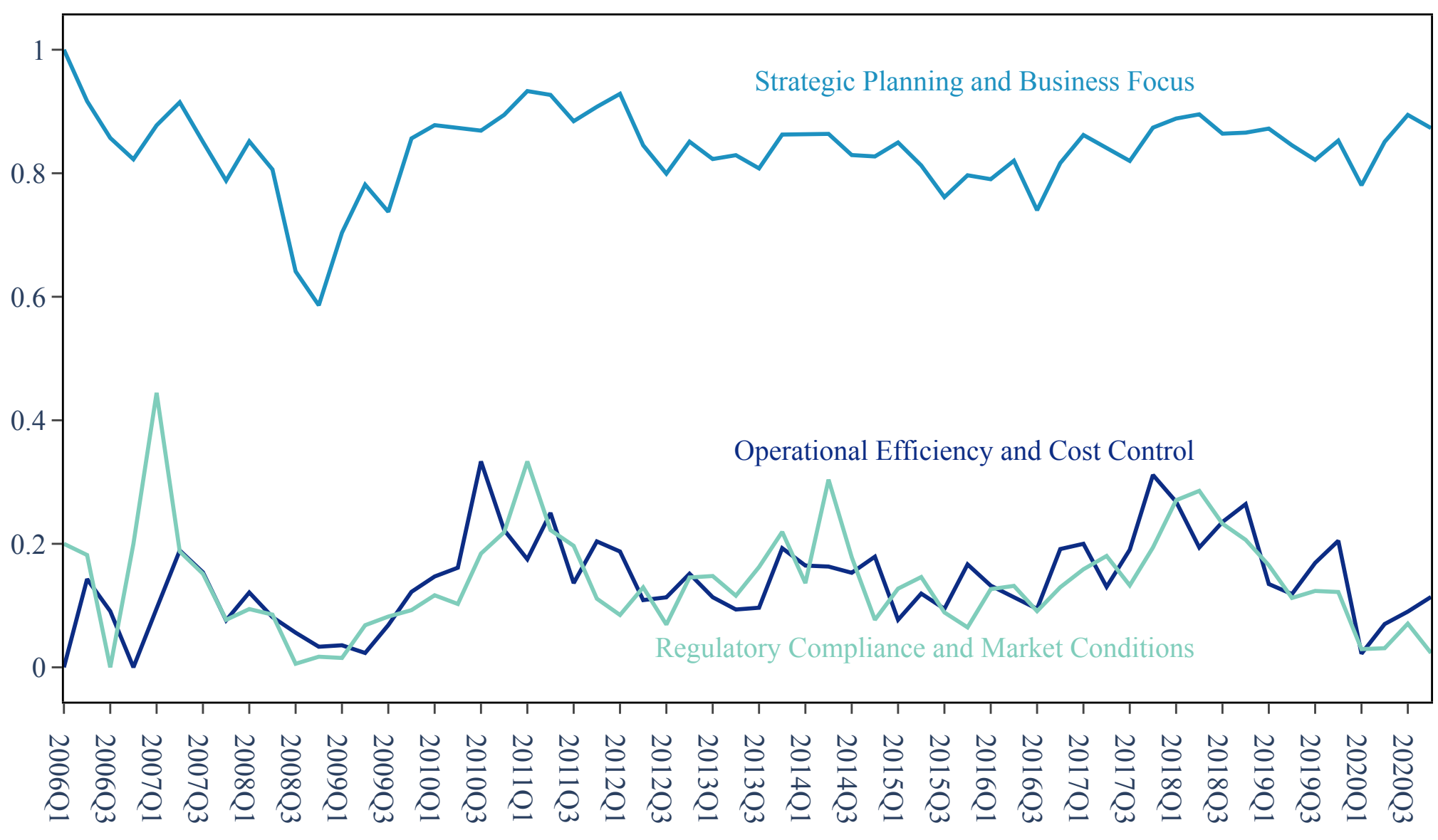

**Table 1.** Summary Statistics

Panel A displays the descriptive statistics of the investment plan derived from earnings call transcripts utilizing ChatGPT (*ChatGPT Investment Score*) and the characteristics of firms participating in the earnings call. Panel B presents the mean firm characteristics and the mean difference for each characteristic across two subsamples based on the ChatGPT investment score. The low (high) subsample is made up of earnings call transcripts with a *ChatGPT Investment Score* less (higher) than zero. The sample comprises Compustat firms with earnings conference call transcripts and non-missing financial variables from 2006 to 2020. Variables are winsorized at 1%. All variables are defined in Appendix A.

Panel A: Descriptive statistics of the whole sample

| | Mean | Median | SD | P25 | P75 | N |
|---|---|---|---|---|---|---|
| ChatGPT Investment Score | 0.014 | 0 | 0.184 | 0 | 0.12 | 74,429 |
| ChatGPT Investment Alt. Score | 0.245 | 0 | 0.328 | 0 | 0.5 | 74,429 |
| | | | | | | |
| **Investment Measures** | | | | | | |
| Capital Expenditure (%) | 1.151 | 0.695 | 1.591 | 0.318 | 1.42 | 74,429 |
| Intangible Capital Stock ($M) | 5,306.403 | 838.392 | 14,142.13 | 212.337 | 3,507.007 | 74,429 |
| Physical Capital Stock ($M) | 2,476.443 | 258.134 | 6,645.288 | 39.554 | 1,369.318 | 74,429 |
| Intangible Investment (%) | 1.99 | 1.556 | 1.897 | 0.614 | 2.717 | 74,429 |
| Physical Investment (%) | 1.222 | 0.583 | 2.269 | 0.24 | 1.423 | 74,429 |
| Total Investment (%) | 3.294 | 2.647 | 2.943 | 1.676 | 4.029 | 74,429 |
| R&D (%) | 1.628 | 1.055 | 1.95 | 0.268 | 2.19 | 74,429 |
| | | | | | | |
| **Return Measures** | | | | | | |
| Return (Annualized, %) | 14.388 | 10.7 | 100.903 | -39.801 | 60.079 | 74,429 |
| FF5-adjusted Return (%) | -2.182 | -1.55 | 94.431 | -49.15 | 43.024 | 74,429 |
| $q$5-adjusted Return (%) | 0.77 | 0.091 | 94.997 | -47.245 | 45.483 | 74,429 |
| | | | | | | |
| **Controls** | | | | | | |
| Total $q$ | 1.164 | 0.853 | 1.044 | 0.471 | 1.465 | 74,429 |
| Total Cash Flow | 0.033 | 0.035 | 0.064 | 0.011 | 0.065 | 74,429 |
| Leverage | 0.229 | 0.174 | 0.219 | 0.046 | 0.35 | 74,429 |
| Book Assets ($M) | 8,341.48 | 1,483.3 | 25,126.64 | 346.87 | 5,570.47 | 74,429 |

Panel B: Comparison between firms with low and high ChatGPT Investment Score

| Variables | Low Score | High Score | Difference | $t$-stat. |
|---|---|---|---|---|
| Capital Expenditure (%) | 1.13 | 1.39 | -0.26 | -15.87*** |
| Intangible Capital Stock ($M) | 5,378.15 | 5,846.93 | -468.77 | -3.28*** |
| Physical Capital Stock ($M) | 2,934.21 | 2,802.25 | 131.97 | 1.88* |
| Intangible Investment (%) | 1.63 | 2.05 | -0.42 | -24.23*** |
| Physical Investment (%) | 1.15 | 1.51 | -0.36 | -14.49*** |
| Total Investment (%) | 2.81 | 3.66 | -0.85 | -27.84*** |
| R&D (%) | 1.14 | 1.62 | -0.48 | -20.66*** |
| Return (Annualized, %) | 19.94 | 12.95 | 6.99 | 6.99*** |
| FF5-adjusted Return (%) | -1.18 | -2.12 | 0.93 | 1.01 |
| $q$5-adjusted Return (%) | 2.15 | 0.82 | 1.33 | 1.43 |
| Total $q$ | 0.85 | 1.41 | -0.56 | -57.23*** |
| Total Cash Flow | 0.02 | 0.04 | -0.02 | -32.09*** |
| Leverage | 0.3 | 0.2 | 0.1 | 47.38*** |
| Book Assets ($M) | 9,078.25 | 9,222.92 | -144.68 | -0.55 |
| Number of observations | 18,548 | 23,000 | | |

**Table 2.** ChatGPT Predictions vs. CFO Survey Results

This table presents coefficients from a firm-quarter level estimation that regresses the Duke CFO Survey-based measure on the ChatGPT predicted measure of corporate capital expenditure in the next 12 months. *ChatGPT Investment Score* measures the capital expenditure change predicted by ChatGPT from firms' earnings call transcripts. *CFO Survey Investment* is the expected capital expenditure change for the next year mentioned by corporate executives during the CFO survey conducted by Duke University. Variables are defined in Appendix A. In all panels, the $t$-statistics, in parentheses, are based on standard errors clustered by industry. ***, **, * denote statistical significance at the 0.01, 0.05, and 0.10 levels, respectively.

| | (1) | (2) |
|---|---|---|
| | *CFO Survey Investment* | |
| *ChatGPT Investment Score* | 30.83*** | 21.78*** |
| | (4.36) | (3.57) |
| Industry FE | N | Y |
| YearQtr FE | N | Y |
| R-squared | 0.014 | 0.070 |
| N | 1,338 | 1,325 |

**Table 3.** ChatGPT Investment Score, Tobin's $q$, and Future Investment

This table reports coefficients from a firm-quarter level estimation that regresses firms' real capital expenditure for the next quarter on the predicted capital expenditure by ChatGPT using earnings call transcripts. *ChatGPT Investment Score* measures the capital expenditure change predicted by ChatGPT from firms' earnings call transcripts. The dependent variable, *Capital Expenditure*, is the real capital expenditure scaled by book assets for quarter $t+2$. Control variables include *Total q* (Peters and Taylor, 2017), *Capital Expenditure*, *Total Cash Flow*, *Market Leverage*, and *Firm Size* in quarter $t$. All variables are defined in Appendix A. In all panels, the $t$-statistics, in parentheses, are based on standard errors clustered by firm. ***, **, * denote statistical significance at the 0.01, 0.05, and 0.10 levels, respectively.

| | (1) | (2) | (3) | (4) |
|---|---|---|---|---|
| | *Capital Expenditure*$_{t+2}$ | | | |
| *ChatGPT Investment Score*$_t$ | 0.682*** | 0.581*** | 0.505*** | 0.477*** |
| | (16.97) | (15.26) | (14.49) | (13.91) |
| *Total q*$_t$ | | 0.223*** | | 0.108*** |
| | | (13.23) | | (7.47) |
| *Capital Expenditure*$_t$ | | | 0.150*** | 0.149*** |
| | | | (8.94) | (8.91) |
| *Total Cash Flow*$_t$ | | | 0.825*** | 0.607*** |
| | | | (4.88) | (3.54) |
| *Leverage*$_t$ | | | -1.461*** | -1.302*** |
| | | | (-14.33) | (-12.93) |
| *Size*$_t$ | | | -0.0961*** | -0.0974*** |
| | | | (-4.13) | (-4.22) |
| Firm FE | Y | Y | Y | Y |
| YearQtr FE | Y | Y | Y | Y |
| R-squared | 0.550 | 0.555 | 0.574 | 0.575 |
| Observations | 74,429 | 74,429 | 74,429 | 74,429 |

**Table 4.** ChatGPT Investment Score and Long-Term Investment

This table presents coefficients from a firm-quarter level estimation that regresses firms' real capital expenditure in subsequent quarters on the predicted capital expenditure by ChatGPT using earnings call transcripts. *ChatGPT Investment Score* measures the capital expenditure change predicted by ChatGPT from firms' earnings call transcripts. The dependent variable, *Capital Expenditure*, is the real capital expenditure scaled by book assets for quarter $t+n$. Control variables include *Total q*, *Capital Expenditure*, *Total Cash Flow*, *Market Leverage*, and *Firm Size* in quarter $t$. All variables are defined in Appendix A. In all panels, the $t$-statistics, in parentheses, are based on standard errors clustered by firm. ***, **, * denote statistical significance at the 0.01, 0.05, and 0.10 levels, respectively.

| | (1) | (2) | (3) | (4) | (5) | (6) | (7) | (8) |
|---|---|---|---|---|---|---|---|---|
| | (n=3) | (n=4) | (n=5) | (n=6) | (n=7) | (n=8) | (n=9) | (n=10) |
| | *Capital Expenditure*$_{t+n}$ | | | | | | | |
| *ChatGPT Investment Score*$_t$ | 0.519*** | 0.474*** | 0.370*** | 0.293*** | 0.231*** | 0.194*** | 0.133*** | 0.0918*** |
| | (14.04) | (13.99) | (11.29) | (8.18) | (7.41) | (6.78) | (4.52) | (3.11) |
| *Total q*$_t$ | 0.131*** | 0.122*** | 0.127*** | 0.135*** | 0.116*** | 0.0957*** | 0.0846*** | 0.0606*** |
| | (8.29) | (7.43) | (8.59) | (8.52) | (7.30) | (6.35) | (6.00) | (4.17) |
| *Capital Expenditure*$_t$ | 0.139*** | 0.0677*** | 0.0102 | 0.00652 | 0.0231 | 0.000937 | -0.0405** | -0.0177 |
| | (7.61) | (4.83) | (0.48) | (0.48) | (1.24) | (0.06) | (-2.02) | (-1.16) |
| *Total Cash Flow*$_t$ | 0.241 | 0.573*** | 0.0204 | -0.203 | -0.329 | 0.116 | 0.0874 | 0.126 |
| | (1.40) | (3.29) | (0.11) | (-1.04) | (-1.61) | (0.61) | (0.47) | (0.68) |
| *Leverage*$_t$ | -1.155*** | -1.022*** | -1.016*** | -0.799*** | -0.624*** | -0.448*** | -0.400*** | -0.301*** |
| | (-11.47) | (-9.89) | (-9.89) | (-8.27) | (-6.32) | (-4.85) | (-4.30) | (-3.18) |
| *Size*$_t$ | -0.128*** | -0.171*** | -0.157*** | -0.157*** | -0.165*** | -0.177*** | -0.173*** | -0.145*** |
| | (-5.05) | (-6.06) | (-6.11) | (-5.57) | (-5.96) | (-6.43) | (-6.29) | (-4.90) |
| Firm FE | Y | Y | Y | Y | Y | Y | Y | Y |
| YearQtr FE | Y | Y | Y | Y | Y | Y | Y | Y |
| R-squared | 0.555 | 0.543 | 0.547 | 0.537 | 0.535 | 0.536 | 0.544 | 0.532 |
| Observations | 74,108 | 73,598 | 72,871 | 71,835 | 70,814 | 69,819 | 68,851 | 67,873 |

**Table 5.** ChatGPT Investment Score, Tobin's q, and Various types of Investment

This table presents coefficients from a firm-quarter level estimation that regresses firms' investment in the subsequent year on the predicted capital expenditure by ChatGPT. *ChatGPT Investment Score* measures the capital expenditure change predicted by ChatGPT from firms' earnings call transcripts. We define *Total q* and various investment variables following Peters and Taylor (2017): *Intangible Capital*, calculated from accumulating R&D and a proportion of SG&A expenses; *Physical Capital*, the PP&E; *Total Capital*, the sum of *Physical capital* and *Intangible capital*; *Total q*, the ratio of market capitalization to *Total Capital*; *Physical Investment*, capital expenditure scaled by *Total Capital*; *Intangible Investment*, R&D + 0.3 × SG&A expenses, scaled by *Total Capital*; *TotalInvestment*, the sum of *Physical investment* and *Intangible investment*. Control variables include *Total q*, *Total Cash Flow*, *Market Leverage*, *Firm Size* dependent variables in quarter $t$. All variables are defined in Appendix A. In all panels, the $t$-statistics, in parentheses, are based on standard errors clustered by firm. ***, **, * denote statistical significance at the 0.01, 0.05, and 0.10 levels, respectively.

| | (1) | (2) | (3) | (4) | (5) | (6) | (7) | (8) |
|---|---|---|---|---|---|---|---|---|
| | *Physical Investment*$_{t+2}$ | | *Intangible Investment*$_{t+2}$ | | *Total Investment*$_{t+2}$ | | *R&D*$_{t+2}$ | |
| *ChatGPT Investment Score*$_t$ | 0.732*** | 0.481*** | 0.274*** | 0.103*** | 1.059*** | 0.621*** | 0.291*** | 0.135*** |
| | (19.98) | (15.55) | (11.93) | (5.99) | (21.78) | (15.44) | (8.40) | (5.37) |
| *Total q*$_t$ | | 0.185*** | | 0.224*** | | 0.513*** | | 0.204*** |
| | | (12.71) | | (15.91) | | (20.80) | | (12.39) |
| *Physical Investment*$_t$ | | 0.239*** | | | | | | |
| | | (17.94) | | | | | | |
| *Intangible Investment*$_t$ & | | | | 0.458*** | | | | |
| | | | | (24.71) | | | | |
| *Total Investment*$_t$ | | | | | | 0.268*** | | |
| | | | | | | (25.31) | | |
| *R&D*$_t$ | | | | | | | | 0.492*** |
| | | | | | | | | (24.82) |
| Control Variables | Y | Y | Y | Y | Y | Y | Y | Y |
| Firm FE | Y | Y | Y | Y | Y | Y | Y | Y |
| YearQtr FE | Y | Y | Y | Y | Y | Y | Y | Y |
| R-squared | 0.666 | 0.710 | 0.857 | 0.899 | 0.686 | 0.746 | 0.853 | 0.905 |
| Observations | 67,621 | 67,621 | 67,621 | 67,621 | 67,621 | 67,621 | 36,303 | 34,253 |

**Table 6.** ChatGPT Investment Score, Tobin's q, and Future Investment: Bias-corrected Estimates

This table reports coefficients from a firm-quarter level estimation that regresses firms' real capital expenditure for the next quarter on the predicted capital expenditure by ChatGPT using earnings call transcripts, all estimated using the bias-corrected cumulant estimators as described in Erickson, Jiang, and Whited (2014). *ChatGPT Investment Score* measures the capital expenditure change predicted by ChatGPT from firms' earnings call transcripts. The dependent variable, *Capital Expenditure*, is the real capital expenditure scaled by book assets for quarter $t+2$. Control variables include *Total q* (Peters and Taylor, 2017), *Capital Expenditure*, *Total Cash Flow*, *Market Leverage*, and *Firm Size* in quarter $t$. All variables are defined in Appendix A. In all panels, the $t$-statistics, in parentheses, are based on standard errors clustered by firm. ***, **, * denote statistical significance at the 0.01, 0.05, and 0.10 levels, respectively.

| | (1) | (2) | (3) | (4) | (5) | (6) |
|---|---|---|---|---|---|---|
| | | | *Capital Expenditure*$_{t+2}$ | | | |
| *Total q*$_t$ | 1.992*** | | 0.171*** | 4.588*** | | 0.187*** |
| | (4.79) | | (3.25) | (2.78) | | (2.60) |
| *ChatGPT Investment Score*$_t$ | | 12.24*** | 1.568*** | | 14.14*** | 1.566*** |
| | | (7.83) | (4.77) | | (6.95) | (4.06) |
| *Capital Expenditure*$_t$ | | | | 0.133*** | 0.185*** | 0.158*** |
| | | | | (6.51) | (8.33) | (9.28) |
| *Total Cash Flow*$_t$ | | | | -7.618** | -3.429*** | 0.321 |
| | | | | (-2.36) | (-4.74) | (1.27) |
| *Leverage*$_t$ | | | | 5.424** | 1.640*** | -0.949*** |
| | | | | (2.16) | (3.37) | (-5.29) |
| *Size*$_t$ | | | | 0.0660 | 0.0544** | 0.0024 |
| | | | | (0.93) | (2.16) | (0.17) |
| Firm FE | Y | Y | Y | Y | Y | Y |
| YearQtr FE | Y | Y | Y | Y | Y | Y |
| Rho-squared | 0.106 | 0.193 | 0.033 | 0.149 | 0.216 | 0.080 |
| Observations | 74,429 | 74,429 | 74,429 | 74,429 | 74,429 | 74,429 |

**Table 7.** ChatGPT Investment Score and Information Environment

This table presents coefficients from a firm-quarter level estimation that regresses firms' total investment in subsequent quarters on the interactions of the ChatGPT investment score and information environment proxies. *ChatGPT Investment Score* measures the capital expenditure change predicted by ChatGPT from firms' earnings call transcripts. The dependent variable, *Total investment*, is the sum of *Physical investment* and *Intangible investment* for quarter $t+2$. Information environment proxies include *HHI*, the Herfindahl–Hirschman Index constructed based on the industry classification of Hoberg and Phillips (2016) for quarter $t$; *Top4Shares*, the sum of market shares of the top 4 firms in an industry for quarter $t$; *Size*, the natural logarithm of total book assets for quarter $t$; and *Life1* to *Life4*, the product life cycle stage measures of Hoberg and Maksimovic (2022). *Life1*-*Life4* represent four stages in the product lifecycle: product innovation, process innovation, stability and maturity, and product discontinuation, respectively. Control variables include *Total q* defined in Peters and Taylor (2017), *Capital Expenditure*, *Total Cash Flow*, and *Market Leverage*, for quarter $t$. All variables are defined in Appendix A. In all panels, the $t$-statistics, in parentheses, are based on standard errors clustered by firm. ***, **, * denote statistical significance at the 0.01, 0.05, and 0.10 levels, respectively.

| | (1) | (2) | (3) | (4) | (5) | (6) |
|---|---|---|---|---|---|---|
| | | | *Total Investment*$_{t+2}$ | | | |
| *ChatGPT Investment Score*$_t$ | 0.794*** | 1.154*** | 1.190*** | 1.741*** | | |
| | (12.41) | (10.16) | (8.16) | (9.11) | | |
| *ChatGPT Investment Score*$_t$ × *HHI*$_t$ | -0.653*** | | | -0.586*** | -0.448*** | -0.303** |
| | (-4.83) | | | (-4.28) | (-3.31) | (-2.31) |
| *ChatGPT Investment Score*$_t$ × *Top4Shares*$_t$ | | -0.898*** | | -0.655*** | -0.747*** | -0.610*** |
| | | (-5.41) | | (-3.89) | (-4.37) | (-3.64) |
| *ChatGPT Investment Score*$_t$ × *Size*$_t$ | | | -0.0795*** | -0.0811*** | -0.0932*** | -0.0656*** |
| | | | (-4.22) | (-4.30) | (-4.62) | (-3.33) |
| *ChatGPT Investment Score*$_t$ × *Life1*$_t$ | | | | | 1.977*** | 1.446*** |
| | | | | | (5.58) | (4.30) |
| *ChatGPT Investment Score*$_t$ × *Life2*$_t$ | | | | | 2.962*** | 2.416*** |
| | | | | | (8.84) | (7.40) |
| *ChatGPT Investment Score*$_t$ × *Life3*$_t$ | | | | | 0.107 | 0.314 |
| | | | | | (0.25) | (0.74) |
| *ChatGPT Investment Score*$_t$ × *Life4*$_t$ | | | | | 0.573 | 0.492 |
| Interactions with *Total q*$_t$ | N | N | N | N | N | Y |
| Control Variables | Y | Y | Y | Y | Y | Y |
| Firm FE | Y | Y | Y | Y | Y | Y |
| YearQtr FE | Y | Y | Y | Y | Y | Y |
| R-squared | 0.751 | 0.745 | 0.745 | 0.751 | 0.752 | 0.754 |
| Observations | 62,971 | 66,515 | 66,515 | 62,971 | 62,767 | 62,767 |

**Table 8.** ChatGPT Investment Score and Future Returns

This table reports coefficients from a firm-quarter level estimation that regresses firms' stock performance in the next quarter on the predicted capital expenditure by ChatGPT using earnings call transcripts. *ChatGPT Investment Score* measures the capital expenditure change predicted by ChatGPT from firms' earnings call transcripts. The dependent variable, *Stock Performance*, is one of the following three measures in quarter $t+2$: Annualized Quarterly Raw Return (*Return*); Annualized Quarterly Fama-French 5-factor alpha (*FF5-Adjusted Return*); Annualized Quarterly $q$-factor alpha (*q5-Adjusted Return*). Control variables include *Total q* and *Return* in quarter $t$. All variables are defined in Appendix A. In all panels, the $t$-statistics, in parentheses, are based on standard errors clustered by firm. ***, **, * denote statistical significance at the 0.01, 0.05, and 0.10 levels, respectively.

| | (1) | (2) | (3) | (4) | (5) | (6) |
|---|---|---|---|---|---|---|
| | $Return_{t+2}$ | | $FF5\text{-}Adjusted\ Return_{t+2}$ | | $q5\text{-}Adjusted\ Return_{t+2}$ | |
| $ChatGPT\ Investment\ Score_t$ | -17.91*** | -9.944*** | -16.19*** | -8.082*** | -14.89*** | -7.731*** |
| | (-8.40) | (-4.57) | (-7.19) | (-3.53) | (-6.70) | (-3.42) |
| $Total\ q_t$ | | -15.67*** | | -13.12*** | | -12.70*** |
| | | (-19.54) | | (-15.85) | | (-14.97) |
| $Return_t$ | | -0.0622*** | | -0.158*** | | -0.102*** |
| | | (-3.07) | | (-7.29) | | (-4.67) |
| Firm FE | Y | Y | Y | Y | Y | Y |
| YearQtr FE | Y | Y | Y | Y | Y | Y |
| R-squared | 0.233 | 0.239 | 0.086 | 0.094 | 0.082 | 0.088 |
| Observations | 74,429 | 74,429 | 74,429 | 74,429 | 74,429 | 74,429 |

**Table 9.** ChatGPT Investment Score and Long-Term Returns

This table reports coefficients from a firm-quarter level estimation that regresses firms' stock performance in subsequent quarters on the predicted capital expenditure by ChatGPT using earnings call transcripts. *ChatGPT Investment Score* measures the capital expenditure change predicted by ChatGPT from firms' earnings call transcripts. The dependent variable, *Stock Performance*, is one of the following three measures in quarter $t+n$: Annualized Quarterly Raw Return (*Return*); Annualized Quarterly Fama-French 5-factor alpha (*FF5-Adjusted Return*); Annualized Quarterly $q$-factor alpha (*q5-Adjusted Return*). Control variables include *Total q* and *Return* in quarter $t$. Variables are defined in Appendix A. In all panels, the $t$-statistics, in parentheses, are based on standard errors clustered by firm. ***, **, * denote statistical significance at the 0.01, 0.05, and 0.10 levels, respectively.

Panel A: ChatGPT Investment Score and Long-Term Raw Return

| | (1) | (2) | (3) | (4) | (5) | (6) | (7) | (8) |
|---|---|---|---|---|---|---|---|---|
| | n=3 | n=4 | n=5 | n=6 | n=7 | n=8 | n=9 | n=10 |
| | $Return_{t+n}$ | | | | | | | |
| $ChatGPT\ Investment\ Score_t$ | -11.63*** | -14.17*** | -9.086*** | -5.914*** | -8.403*** | -3.049 | -6.443*** | -2.980 |
| | (-5.39) | (-6.62) | (-4.31) | (-2.60) | (-3.92) | (-1.39) | (-2.88) | (-1.29) |
| $Total\ q_t$ | -13.29*** | -9.240*** | -9.822*** | -8.795*** | -8.038*** | -8.362*** | -6.716*** | -5.594*** |
| | (-17.64) | (-12.55) | (-12.89) | (-12.01) | (-10.76) | (-11.14) | (-8.74) | (-6.93) |
| $Return_t$ | -0.0165*** | -0.0668*** | 0.0111** | -0.0206*** | 0.0000600 | -0.0283*** | -0.0281*** | 0.00569 |
| | (-3.17) | (-13.20) | (2.08) | (-3.90) | (0.01) | (-5.08) | (-4.85) | (0.93) |
| Firm FE | Y | Y | Y | Y | Y | Y | Y | Y |
| YearQtr FE | Y | Y | Y | Y | Y | Y | Y | Y |
| R-squared | 0.227 | 0.239 | 0.225 | 0.226 | 0.227 | 0.225 | 0.228 | 0.224 |
| N | 73,437 | 72,354 | 71,003 | 68,215 | 65,393 | 63,267 | 60,437 | 57,799 |

Panel B: ChatGPT Investment Score and FF5-Adjusted Alpha Raw Return

| | (1) | (2) | (3) | (4) | (5) | (6) | (7) | (8) |
|---|---|---|---|---|---|---|---|---|
| | n=3 | n=4 | n=5 | n=6 | n=7 | n=8 | n=9 | n=10 |
| | *FF-5 factor Adjusted* $Alpha_{t+n}$ | | | | | | | |
| $ChatGPT\ Investment\ Score_t$ | -5.528** | -3.889* | -5.946*** | -6.648*** | -2.218 | -1.025 | -5.970** | -4.863** |
| | (-2.50) | (-1.73) | (-2.71) | (-2.92) | (-0.99) | (-0.45) | (-2.53) | (-2.09) |
| $Total\ q_t$ | -11.59*** | -10.51*** | -8.728*** | -7.089*** | -6.911*** | -7.679*** | -7.273*** | -6.323*** |
| | (-14.55) | (-13.45) | (-10.67) | (-8.85) | (-8.52) | (-9.07) | (-8.72) | (-7.17) |
| $Return_t$ | -0.0235*** | -0.0376*** | -0.0132** | -0.0294*** | -0.00203 | 0.0148** | 0.00189 | 0.00134 |
| | (-4.31) | (-6.85) | (-2.37) | (-5.18) | (-0.36) | (2.41) | (0.31) | (0.21) |
| Firm FE | Y | Y | Y | Y | Y | Y | Y | Y |
| YearQtr FE | Y | Y | Y | Y | Y | Y | Y | Y |
| R-squared | 0.0867 | 0.0917 | 0.0896 | 0.0906 | 0.0892 | 0.0928 | 0.0967 | 0.0911 |
| N | 73,437 | 72,354 | 71,003 | 68,215 | 65,393 | 63,267 | 60,437 | 57,799 |

Panel C: ChatGPT Investment Score and $q5$-Adjusted Alpha Raw Return

| | (1) | (2) | (3) | (4) | (5) | (6) | (7) | (8) |
|---|---|---|---|---|---|---|---|---|
| | n=3 | n=4 | n=5 | n=6 | n=7 | n=8 | n=9 | n=10 |
| | *q5-Adjusted* $Alpha_{t+n}$ | | | | | | | |
| $ChatGPT\ Investment\ Score_t$ | -8.329*** | -9.343*** | -8.413*** | -9.722*** | -8.764*** | -8.316*** | -9.012*** | -5.977** |
| | (-3.74) | (-4.22) | (-3.84) | (-4.20) | (-3.98) | (-3.62) | (-3.78) | (-2.47) |
| $Total\ q_t$ | -9.640*** | -8.606*** | -8.819*** | -7.923*** | -8.648*** | -9.215*** | -8.237*** | -6.830*** |
| | (-11.72) | (-10.43) | (-10.43) | (-9.53) | (-10.22) | (-10.60) | (-9.08) | (-7.31) |
| $Return_t$ | -0.0460*** | -0.0282*** | 0.00228 | -0.00958* | -0.00314 | -0.00683 | -0.0167*** | 0.0119* |
| | (-8.31) | (-5.07) | (0.41) | (-1.68) | (-0.55) | (-1.11) | (-2.72) | (1.89) |
| Firm FE | Y | Y | Y | Y | Y | Y | Y | Y |
| YearQtr FE | Y | Y | Y | Y | Y | Y | Y | Y |
| R-squared | 0.0838 | 0.0846 | 0.0829 | 0.0836 | 0.0871 | 0.0875 | 0.0903 | 0.0863 |
| N | 73,437 | 72,354 | 71,003 | 68,215 | 65,393 | 63,267 | 60,437 | 57,799 |

**Table 10.** ChatGPT Investment Score and Short-Term Returns

This table presents coefficients from a firm-quarter level estimation that regresses firms' short-term cumulative abnormal return following the earnings call date on the predicted capital expenditure by ChatGPT. *ChatGPT Investment Score* measures the capital expenditure change predicted by ChatGPT from firms' earnings call transcripts. The dependent variable *CAR[0,1]*, *CAR[0,3]*, and *CAR[0,5]* represents 2-day, 4-day, and 6-day Carhart 4-factor adjusted accumulative abnormal return after the earnings call date for quarter t respectively. Control variables include *Total q*, *Capital Expenditure*, *Total Cash Flow*, *Market Leverage*, *Firm Size*, *Sentiment*, and *Earnings Surprise* in quarter $t$. All variables are defined in Appendix A. In all panels, the $t$-statistics, in parentheses, are based on standard errors clustered by firm. ***, **, * denote statistical significance at the 0.01, 0.05, and 0.10 levels, respectively.

| | (1) | (2) | (3) | (4) | (5) | (6) |
|---|---|---|---|---|---|---|
| | *CAR[0,1]* | | *CAR[0,3]* | | *CAR[0,5]* | |
| *ChatGPT Investment Score*$_t$ | 3.176*** | 3.139*** | 3.162*** | 3.119*** | 3.236*** | 3.181*** |
| | (12.72) | (10.43) | (11.46) | (9.19) | (11.06) | (8.97) |
| *Total q*$_t$ | -1.066*** | -1.061*** | -1.183*** | -1.190*** | -1.282*** | -1.300*** |
| | (-10.88) | (-8.84) | (-10.83) | (-8.98) | (-10.74) | (-9.10) |
| *Return*$_t$ | -0.266*** | -0.262*** | -0.321*** | -0.249*** | -0.287*** | -0.225*** |
| | (-4.76) | (-3.66) | (-4.99) | (-3.20) | (-4.23) | (-2.69) |
| *Total Cash Flow*$_t$ | 9.351*** | 7.297*** | 10.37*** | 8.425*** | 9.780*** | 8.373*** |
| | (7.04) | (4.81) | (7.24) | (4.84) | (6.38) | (4.36) |
| *Leverage*$_t$ | 4.199*** | 3.597*** | 5.340*** | 4.445*** | 6.068*** | 5.154*** |
| | (6.76) | (4.68) | (7.77) | (5.40) | (8.41) | (5.83) |
| *Size*$_t$ | -1.202*** | -1.188*** | -1.389*** | -1.415*** | -1.483*** | -1.487*** |
| | (-9.00) | (-7.29) | (-9.34) | (-7.90) | (-9.30) | (-7.80) |
| *Sentiment*$_t$ | 9.307*** | 9.457*** | 9.520*** | 9.631*** | 9.479*** | 9.526*** |
| | (31.19) | (25.33) | (29.12) | (23.21) | (26.98) | (21.60) |
| *Earnings Surprise*$_t$ | | 0.281* | | 0.355** | | 0.173 |
| | | (1.71) | | (2.13) | | (0.88) |
| Firm FE | Y | Y | Y | Y | Y | Y |
| YearQtr FE | Y | Y | Y | Y | Y | Y |
| R-squared | 0.109 | 0.112 | 0.106 | 0.109 | 0.102 | 0.105 |
| N | 73,542 | 43,103 | 73,542 | 43,103 | 73,542 | 43,103 |

**Table 11.** ChatGPT Investment Score and Changes in Analyst Forecasts

This table presents coefficients from a firm-quarter level estimation that regresses the change in analysts' capital expenditure forecast for the next quarter around the date of the earnings call on the predicted capital expenditure by ChatGPT using earnings call transcripts. *ChatGPT Investment Score* measures the capital expenditure change predicted by ChatGPT from firms' earnings call transcripts. The dependent variable, *Change in Analyst Forecast*, is the change in analysts' capital expenditure forecast for quarter $t+1$ around the date of the earnings call scaled by capital expenditure in quarter $t$, multiplied by 100. Control variables include *Total q* (Peters and Taylor, 2017), *Capital Expenditure*, *Total Cash Flow*, *Market Leverage*, and *Firm Size* in quarter $t$. All variables are defined in Appendix A. In all panels, the $t$-statistics, in parentheses, are based on standard errors clustered by firm. ***, **, * denote statistical significance at the 0.01, 0.05, and 0.10 levels, respectively.

| | (1) | (2) | (3) | (4) |
|---|---|---|---|---|
| | *Change in Analyst Forecast*$_{t+1}$ | | | |
| *ChatGPT Investment Score*$_t$ | 8.990*** | 8.513*** | 8.205*** | 7.952*** |
| | (12.68) | (11.96) | (11.40) | (11.06) |
| *Total q*$_t$ | | 0.641*** | | 0.532*** |
| | | (3.66) | | (2.88) |
| *Capital Expenditure*$_t$ | | | 0.0393 | 0.0276 |
| | | | (0.23) | (0.16) |
| *Total Cash Flow*$_t$ | | | -0.947 | -2.269 |
| | | | (-0.45) | (-1.06) |
| *Leverage*$_t$ | | | -8.709*** | -7.843*** |
| | | | (-5.26) | (-4.70) |
| *Size*$_t$ | | | 0.404 | 0.508 |
| | | | (1.00) | (1.27) |
| YearQtr FE | Y | Y | Y | Y |
| Firm FE | Y | Y | Y | Y |
| R-squared | 0.112 | 0.113 | 0.113 | 0.114 |
| Observations | 34,080 | 34,080 | 34,080 | 34,080 |

**Table 12.** Factors that Drive Expected Investment

This table presents coefficients from a firm-quarter level estimation that regresses firms' real capital expenditure in subsequent quarters on the interactions of the ChatGPT Investment Score with dummy variables representing the reasons provided by ChatGPT for its response. *Operational Efficiency*, *Strategic Planning*, and *External Conditions* represent three broad categories: (i) Operational Efficiency and Cost Control, (ii) Strategic Planning and Business Focus, and (iii) Regulatory Compliance and Market Conditions, classified based on the one-sentence explanations provided by ChatGPT for its investment assessments. *ChatGPT Inv. Score* measures the capital expenditure change predicted by ChatGPT from firms' earnings call transcripts. The dependent variable, *Capital Expenditure*, is the real capital expenditure scaled by book assets for quarter $t+n$. Control variables include *Total q*, *Capital Expenditure*, *Total Cash Flow*, *Market Leverage*, and *Firm Size* in quarter $t$. All variables are defined in Appendix A. In all panels, the $t$-statistics, in parentheses, are based on standard errors clustered by firm. ***, **, * denote statistical significance at the 0.01, 0.05, and 0.10 levels, respectively.

| | (1) | (2) | (3) | (4) | (5) | (6) | (7) | (8) |
|---|---|---|---|---|---|---|---|---|
| | (n=2) | (n=3) | (n=4) | (n=5) | (n=6) | (n=7) | (n=8) | (n=9) |
| | *Capital Expenditure*$_{t+n}$ | | | | | | | |
| *ChatGPT Inv. Score*$_t$ × *Oper. Efficiency* | 0.492*** | 0.404*** | 0.366*** | 0.334*** | 0.223*** | 0.168*** | 0.145*** | 0.160*** |
| | (7.65) | (6.16) | (5.89) | (6.24) | (4.07) | (2.66) | (2.70) | (3.26) |
| *ChatGPT Inv. Score*$_t$ × *Strategic Planning* | 0.480*** | 0.529*** | 0.482*** | 0.398*** | 0.334*** | 0.255*** | 0.256*** | 0.157*** |
| | (11.24) | (11.65) | (10.83) | (10.06) | (6.82) | (6.00) | (6.73) | (3.69) |
| *ChatGPT Inv. Score*$_t$ × *External Conditions* | 0.521*** | 0.711*** | 0.671*** | 0.387*** | 0.296*** | 0.282*** | 0.0592 | 0.0451 |
| | (6.96) | (10.00) | (9.69) | (4.96) | (3.60) | (3.39) | (0.68) | (0.63) |
| *Total q*$_t$ | 0.109*** | 0.132*** | 0.122*** | 0.127*** | 0.134*** | 0.117*** | 0.0944*** | 0.0847*** |
| | (7.47) | (8.29) | (7.45) | (8.56) | (8.50) | (7.33) | (6.24) | (5.97) |
| *Capital Expenditure*$_t$ | 0.145*** | 0.133*** | 0.0673*** | 0.0105 | 0.0177 | 0.0210 | -0.00287 | -0.0386* |
| | (8.29) | (7.14) | (4.88) | (0.50) | (1.24) | (1.14) | (-0.18) | (-1.95) |
| *Total Cash Flow*$_t$ | 0.624*** | 0.262 | 0.590*** | 0.0233 | -0.241 | -0.251 | 0.116 | 0.0894 |
| | (3.59) | (1.52) | (3.38) | (0.12) | (-1.22) | (-1.25) | (0.61) | (0.48) |
| *Leverage*$_t$ | -1.303*** | -1.162*** | -1.018*** | -1.017*** | -0.794*** | -0.618*** | -0.458*** | -0.404*** |
| | (-12.94) | (-11.52) | (-9.85) | (-9.92) | (-8.24) | (-6.31) | (-4.96) | (-4.34) |
| *Size*$_t$ | -0.0965*** | -0.127*** | -0.171*** | -0.157*** | -0.157*** | -0.167*** | -0.177*** | -0.173*** |
| | (-4.18) | (-5.01) | (-6.05) | (-6.12) | (-5.59) | (-5.97) | (-6.43) | (-6.32) |
| YearQtrFE | Yes | Yes | Yes | Yes | Yes | Yes | Yes | Yes |
| FirmFE | Yes | Yes | Yes | Yes | Yes | Yes | Yes | Yes |
| R-squared | 0.575 | 0.555 | 0.543 | 0.548 | 0.537 | 0.535 | 0.536 | 0.545 |
| N | 74,429 | 74,110 | 73,596 | 72,870 | 71,831 | 70,817 | 69,817 | 68,851 |

**Table 13.** ChatGPT and Other Corporate Policies: Dividends and Employment

This table presents coefficients from a firm-quarter level estimation that regresses the Duke CFO Survey-based measure with the ChatGPT predicted measure for other corporate policies. The dependent variable, *CFO Survey Dividend* or the *CFO Survey Employment*, is the expected change in dividend payout or the number of employees for the next year mentioned by corporate executives in the Duke CFO survey. *ChatGPT Dividend Score* or *ChatGPT Employment Score* measures the dividend payout or the number of employees derived from firms' earnings call transcripts by ChatGPT of the same quarter. All variables are defined in Appendix A. In all panels, the $t$-statistics, in parentheses, are based on standard errors clustered by industry. ***, **, * denote statistical significance at the 0.01, 0.05, and 0.10 levels, respectively.

| | (1) | (2) | (3) | (4) |
|---|---|---|---|---|
| | *CFO Survey Dividend* | | *CFO Survey Employment* | |
| *ChatGPT Dividend Score* | 45.62*** | 30.46*** | | |
| | (3.99) | (3.93) | | |
| *ChatGPT Employment Score* | | | 22.64*** | 18.01*** |
| | | | (3.00) | (5.20) |
| Industry FE | N | Y | N | Y |
| YearQtr FE | N | Y | N | Y |
| R-squared | 0.023 | 0.117 | 0.007 | 0.044 |
| N | 666 | 661 | 1,322 | 1,311 |

## Internet Appendix to "ChatGPT and Corporate Policies"

## Appendix A: Definitions of Variables

| Variable | Definition |
|---|---|
| *Capital Expenditure* | Capital Expenditure (CAPX) at the end of the quarter, scaled by book assets. |
| *CAR[0,1], CAR[0,3], and CAR[0,5]* | Cumulative abnormal returns following the earnings call date for the windows [0,1], [0,3], and [0,5], estimated using the Carhart 4-factor model. |
| *CFO Survey Dividend* | Executives' response about the firm's Dividend payout plan for the next year during the Duke CFO survey. It is a percentage change compared to the dividend payout in the past 12 months. |
| *CFO Survey Employment* | Executives' response about the firm's Dividend payout plan for the next year during the Duke CFO survey. It is a percentage change compared to the number of employees in the past 12 months. |
| *CFO Survey Investment* | Executives' response about the firm's capital expenditure plan for the next year during the Duke CFO survey. It is a percentage change compared to the capital expenditure in the past 12 months. |
| *Change in Analyst Forecast* | The change in analysts' consensus forecast for capital expenditure for quarter t + 1, as measured around the date of the earnings call, and scaled by the capital expenditure in quarter t, then multiplied by 100. |
| *ChatGPT Dividend Score* | We ask ChatGPT to provide a response about the firm's dividend payout plan in the next year from chunks of earnings call transcripts. Based on the response from the model, we assign a score of -1, -0.5, 0, 0.5, and 1 for each of the given choices: Substantial Decrease; Decrease; No change; Increase; Substantial Increase. We then take the average of the scores across multiple chunks of one earnings call. |
| *ChatGPT Employee Score* | We ask ChatGPT to provide a response about the firm's number of workforce plan in the next year from chunks of earnings call transcripts. Based on the response from the model, we assign a score of -1, -0.5, 0, 0.5, and 1 for each of the given choices: Substantial Decrease; Decrease; No change; Increase; Substantial Increase. We then take the average of the scores across multiple chunks of one earnings call. |

*(continued)*

| **Variable** | **Definition** |
|---|---|
| *ChatGPT Investment Score* | We ask ChatGPT to provide a response about the firm's capital expenditure plan in the next year from chunks of earnings call transcripts. Based on the response from the model, we assign a score of -1, -0.5, 0, 0.5, and 1 for each of the given choices: Substantial Decrease; Decrease; No change; Increase; Substantial Increase. We then take the average of the scores across multiple chunks of one earnings call. |
| *ChatGPT Investment Alt. Score* | We ask ChatGPT to provide a response about the firm's capital expenditure plan in the next year from chunks of earnings call transcripts. Based on the response from the model, we assign a score of -1, -0.5, 0, 0.5, and 1 for each of the given choices: Substantial Decrease; Decrease; No change; Increase; Substantial Increase. We then take the score from the chunk with the largest absolute value across multiple chunks of one earnings call as the final score for that earnings call. |
| *Earnings Surprise* | The change in Earnings Per Share (EPS) from quarter $t-4$ and quarter $t$, divided by the stock price in quarter $t$, following Livnat and Mendenhall (2006). |
| *FF5-Adjusted Return* | Average monthly Fama-French 5-factor abnormal return over one quarter multiplied by 12. |
| *HHI* | The sum of squared market shares in the industry based on textual similarity of firms' 10K product descriptions (Hoberg and Phillips, 2016). |
| *Intangible Capital Stock* | Based on the measure of annual *Intangible Capital Stock* proposed by Peters and Taylor (2017), which applies the perpetual inventory method to firms' intangible investments defined as Research and Development (R&D) and 0.3 × selling, general, and administrative (SG&A) spending at the end of the year, we apply the same method to derive a quarterly measure of *Intangible Capital Stock* assuming a 2.5% quarterly depreciation rate. |
| *Intangible Investment* | Research and Development (R&D) and 0.3 × selling, general, and administrative (SG&A) spending at the end of the quarter, scaled by total capital stock. |
| *Leverage* | The sum of long-term debt (*dlttq*) and short-term debt (*dlcq*) divided by the sum of long-term debt and short-term debt plus the market value of equity (cshoq*prccq) at the end of the quarter. |

*(continued)*

| Variable | Definition |
|---|---|
| *Life1, Life2, Life3 and Life4* | Firms' product life cycle stages defined by Hoberg and Maksimovic (2022), who characterizes the stages of a firm's product portfolio as a four-element vector, where each element is bounded between 0 and 1 and the sum of the four components is 1. *Life1*, *Life2*, *Life3* and *Life4* refer to product innovation, process innovation, stability and maturity, and product discontinuation, respectively. |
| *Physical Capital Stock* | Property, Plant and Equipment at the end of the quarter. |
| *Physical Investment* | Capital Expenditure (CAPX) at the end of the quarter, scaled by *Total Capital Stock*. |
| *Profitability* | Earnings before interest and tax at the end of the quarter, scaled by book assets. |
| *q5-Adjusted Return* | Average monthly q5-factor abnormal return over one quarter multiplied by 12. |
| *R&D* | Research and Development (R&D), scaled by *Total Capital Stock*. |
| *Return* | Annualized buy-and-hold returns over one quarter. |
| *RoBERTa Investment Score* | We ask RoBERTa to provide a response about the firm's capital expenditure plan in the next year from 300-word chunks of earnings call transcripts. Based on the response from the model, we assign a score of -1, -0.5, 0, 0.5, and 1 for each of the given choices: Substantial Decrease; Decrease; No change; Increase; Substantial Increase. We then take the average of the scores across multiple chunks of one earnings call. |
| *Sales Growth* | Percentage of sales growth rate at the end of each quarter. |
| *Sentiment* | The number of positive words minus the number of negative words divided by the sum of the number of positive words and negative words where positive words and negatives are classified following Loughran-McDonald Dictionary. |
| *Size* | The natural logarithm of total book assets at the end of the quarter. |
| *Top4Shares* | The sum of the market shares of the top four firms in an industry for a given quarter. |
| *Total Capital Stock* | The sum of *Physical Capital Stock* and *Intangible Capital Stock* at the end of the quarter. |
| *Total Cash Flow* | Divide total capital by the sum of income before extraordinary items plus depreciation expenses plus after-tax intangible investment (the marginal tax rate is assumed to be 30%). |
| *Total Investment* | The sum of *Physical Investment* and *Intangible Investment* at the end of the quarter. |
| *Total q* | The ratio of market capitalization (calculated using Compustat items $prcc_f \times csho$), plus the book value of debt ($dltt + dlc$), minus the firm's current assets ($act$), to *Total capital stock* for the quarter preceding the earnings call date, as defined in Peters and Taylor (2017). |

*(continued)*

| **Variable** | **Definition** |
|---|---|
| *Total* $q_c$ | The ratio of market capitalization on the earnings call date (calculated using price*shares outstanding), plus the book value of debt ($dltt + dlc$), minus the firm's current assets ($act$), to *Total capital stock* as defined in Peters and Taylor (2017). |
| *Total* $q_{c+1}$ | The ratio of market capitalization on the first trading day after the earnings call date (calculated using price*shares outstanding), plus the book value of debt ($dltt + dlc$), minus the firm's current assets ($act$), to *Total capital stock* as defined in Peters and Taylor (2017). |
| *Total* $q_{c+5}$ | The ratio of market capitalization on the fifth trading day after the earnings call date (calculated using price*shares outstanding), plus the book value of debt ($dltt + dlc$), minus the firm's current assets ($act$), to *Total capital stock* as defined in Peters and Taylor (2017). |
| *Z-score* | Calculated as $3.3 \times$ *Operating Income Before Depreciation* + *Sales/Turnover* + $1.4 \times$ *Retained Earnings* + $1.2 \times$ (*Current Assets* − *Current Liabilities*) / *Assets* |

## Appendix B: Examples of Texts with Predicted Investment Scores

| Category | Example Texts from Conference Call Transcripts |
|---|---|
| *Significantly Increase (Score=1)* | "We committed approximately $250 million of incremental growth capital expenditures compared to our previous allocated budget for new projects to accelerate our investments in Safety Products, Intelligrated and other growth opportunities. These are high-return investments expected to generate triple-digit IRRs. Kingdom, India, the UAE and China."<br><br>"Capital expenditures continued to be higher as we provisioned existing orders and built out for SaaS and PaaS growth. As a reminder, our cloud data centers are built using our own engineered systems. So, while CapEx is a cost to other cloud providers, a good portion of our CapEx is essentially a hardware sale which we sell as a cloud subscription."<br><br>"We invested $3.1 billion in capital expenditures, consistent with our plan for accelerated investment, as we added both commercial and consumer global cloud capacity to meet near-term and longer-term customer demand."<br><br>"We have identified several key strategic initiatives for 2015 to sustain the growth rate of our business. We plan to make significant capital investments in our facilities and infrastructure, and we continue to strengthen our human capital in compliance, manufacturing and sales. We also have a solid slate of plan launches throughout the year." |
| *Significantly Decrease (Score=−1)* | "We have significantly lowered our capital spending plans and are aggressively pursuing operating efficiencies and cost savings as we continue to ramp up production from our major projects, all of which will support cash flow moving forward."<br><br>"As mentioned, the optimization plan includes some business and international market exits, all of which had negligible margin. For perspective, these businesses and markets were a drag of about 20 basis points on 2019 revenue growth and about 40 basis points on 2019 margins. We are also lowering our 2020 CapEx forecast by $10 million to incorporate the exit."<br><br>"After next year we will not have that roughly $50 million to $60 million spend that we'll have this year and next year on El Dorado. So, our CapEx will be down substantially, which will affect - that's a boost of $50 million to $60 million."<br><br>"We are transforming our manufacturing footprint in a way that will enable us to improve flexibility and profitability, while also lowering capital expenditures." |

## Appendix C: Additional Empirical Results

**Table IA.1.** ChatGPT Investment Score and Future Investment: Controlling for Sentiment

This table presents coefficients from a firm-quarter level estimation that regresses firms' real capital expenditure in subsequent quarters on the predicted capital expenditure by ChatGPT using earnings call transcripts. *ChatGPT Investment Score* measures the capital expenditure change predicted by ChatGPT from firms' earnings call transcripts. The dependent variable, *Capital Expenditure*, is the real capital expenditure scaled by book assets for quarter $t+n$. Control variables include *Total q*, *Capital Expenditure*, *Total Cash Flow*, *Market Leverage*, *Firm Size*, and *Sentiment* in quarter $t$. All variables are defined in Appendix A. In all panels, the $t$-statistics, in parentheses, are based on standard errors clustered by firm. ***, **, * denote statistical significance at the 0.01, 0.05, and 0.10 levels, respectively.

| | (1) | (2) | (3) | (4) |
|---|---|---|---|---|
| | *Capital Expenditure*$_{t+2}$ | | | |
| *ChatGPT Investment Score*$_t$ | 0.682*** | 0.581*** | 0.513*** | 0.485*** |
| | (16.97) | (15.26) | (14.38) | (13.80) |
| *Total* $q_t$ | | 0.224*** | | 0.109*** |
| | | (13.24) | | (7.49) |
| *Capital Expenditure*$_t$ | | | 0.146*** | 0.145*** |
| | | | (8.31) | (8.28) |
| *Total Cash Flow*$_t$ | | | 0.853*** | 0.634*** |
| | | | (4.94) | (3.63) |
| *Leverage*$_t$ | | | -1.467*** | -1.306*** |
| | | | (-14.36) | (-12.95) |
| *Size*$_t$ | | | -0.0958*** | -0.0971*** |
| | | | (-4.11) | (-4.20) |
| *Sentiment*$_t$ | | | -0.0417 | -0.0427 |
| | | | (-1.17) | (-1.20) |
| Firm FE | Y | Y | Y | Y |
| YearQtr FE | Y | Y | Y | Y |
| R-squared | 0.551 | 0.556 | 0.574 | 0.575 |
| Observations | 74,429 | 74,429 | 74,429 | 74,429 |

**Table IA.2.** ChatGPT Investment Score and Long-Term Investment: Consistent Sample

This table reports coefficients from a firm-quarter level estimation that regresses firms' real capital expenditure in subsequent quarters on the predicted capital expenditure by ChatGPT keeping the sample constant across different quarters. *ChatGPT Investment Score* is the capital expenditure change predicted by ChatGPT from firms' earnings call transcripts in quarter $t$. The dependent variable, *Capital Expenditure*, is the real capital expenditure scaled by book assets for quarter $t+n$. Control variables include *Total q*, *Capital Expenditure*, *Total Cash Flow*, *Market Leverage*, and *Firm Size* in quarter $t$. Variables are defined in Appendix A. In all panels, the $t$-statistics, in parentheses, are based on standard errors clustered by firm. ***, **, * denote statistical significance at the 0.01, 0.05, and 0.10 levels, respectively.

| | (1) | (2) | (3) | (4) | (5) | (6) | (7) | (8) |
|---|---|---|---|---|---|---|---|---|
| | n=3 | n=4 | n=5 | n=6 | n=7 | n=8 | n=9 | n=10 |
| | *Capital Expenditure*$_{t+n}$ | | | | | | | |
| *ChatGPT Investment Score*$_t$ | 0.495*** | 0.477*** | 0.369*** | 0.300*** | 0.226*** | 0.191*** | 0.129*** | 0.0899*** |
| | (13.21) | (13.38) | (11.66) | (8.21) | (7.24) | (6.48) | (4.43) | (3.01) |
| *Total q*$_t$ | 0.125*** | 0.116*** | 0.125*** | 0.133*** | 0.114*** | 0.0930*** | 0.0862*** | 0.0614*** |
| | (8.21) | (7.23) | (8.24) | (8.16) | (7.06) | (6.23) | (6.01) | (4.21) |
| *Capital Expenditure*$_t$ | 0.133*** | 0.0596*** | 0.0107 | 0.0195 | 0.0207 | -0.00186 | -0.0395** | -0.0188 |
| | (6.40) | (4.63) | (0.47) | (1.09) | (1.04) | (-0.12) | (-1.97) | (-1.26) |
| *Total Cash Flow*$_t$ | 0.298 | 0.627*** | 0.0771 | -0.211 | -0.244 | 0.160 | 0.0953 | 0.101 |
| | (1.61) | (3.57) | (0.41) | (-1.02) | (-1.18) | (0.83) | (0.50) | (0.54) |
| *Leverage*$_t$ | -1.149*** | -1.041*** | -0.980*** | -0.757*** | -0.617*** | -0.440*** | -0.405*** | -0.309*** |
| | (-10.80) | (-9.68) | (-9.18) | (-7.46) | (-6.11) | (-4.68) | (-4.35) | (-3.28) |
| *Size*$_t$ | -0.114*** | -0.155*** | -0.151*** | -0.150*** | -0.160*** | -0.171*** | -0.172*** | -0.146*** |
| | (-4.50) | (-5.47) | (-5.75) | (-5.23) | (-5.67) | (-6.21) | (-6.18) | (-4.93) |
| Firm FE | Y | Y | Y | Y | Y | Y | Y | Y |
| YearQtr FE | Y | Y | Y | Y | Y | Y | Y | Y |
| R-squared | 0.560 | 0.549 | 0.554 | 0.541 | 0.540 | 0.541 | 0.544 | 0.532 |
| Observations | 67,589 | 67,589 | 67,589 | 67,589 | 67,589 | 67,589 | 67,589 | 67,589 |

**Table IA.3.** ChatGPT Investment Score and Long-Term Total Investment

This table reports coefficients from a firm-quarter level estimation that regresses firms' real capital expenditure in subsequent quarters on the predicted capital expenditure by ChatGPT using earnings call transcripts. *ChatGPT Investment Score* measures the capital expenditure change predicted by ChatGPT from firms' earnings call transcripts. The dependent variable, *Total investment*, is the sum of *Physical investment* and *Intangible investment* for quarter $t+n$. Control variables include *Total q*, *Capital Expenditure*, *Total Cash Flow*, *Market Leverage*, and *Firm Size* in quarter $t$. Variables are defined in Appendix A. In all panels, the $t$-statistics, in parentheses, are based on standard errors clustered by firm. ***, **, * denote statistical significance at the 0.01, 0.05, and 0.10 levels, respectively.

| | (1) | (2) | (3) | (4) | (5) | (6) | (7) | (8) |
|---|---|---|---|---|---|---|---|---|
| | n=3 | n=4 | n=5 | n=6 | n=7 | n=8 | n=9 | n=10 |
| | *Total Investment*$_{t+n}$ | | | | | | | |
| *ChatGPT Investment Score*$_t$ | 0.655*** | 0.574*** | 0.473*** | 0.437*** | 0.345*** | 0.242*** | 0.181*** | 0.142*** |
| | (16.26) | (14.83) | (12.23) | (11.56) | (9.30) | (6.72) | (4.92) | (3.90) |
| *Total* $q_t$ | 0.491*** | 0.479*** | 0.482*** | 0.449*** | 0.377*** | 0.342*** | 0.337*** | 0.270*** |
| | (19.30) | (17.55) | (17.15) | (15.99) | (13.68) | (12.14) | (11.91) | (9.58) |
| *Total Investment*$_t$ | 0.271*** | 0.149*** | 0.0991*** | 0.0782*** | 0.115*** | 0.0377*** | 0.0108 | 0.0164* |
| | (23.47) | (15.24) | (7.83) | (8.24) | (9.31) | (3.71) | (0.93) | (1.77) |
| *Total Cash Flow*$_t$ | 0.209 | 1.338*** | 0.353 | -0.173 | -0.416 | 0.716*** | -0.112 | -0.199 |
| | (0.93) | (5.27) | (1.39) | (-0.69) | (-1.57) | (2.66) | (-0.43) | (-0.77) |
| *Leverage*$_t$ | -1.191*** | -1.085*** | -0.994*** | -0.849*** | -0.637*** | -0.457*** | -0.387*** | -0.286** |
| | (-10.67) | (-8.96) | (-7.96) | (-7.01) | (-5.51) | (-3.93) | (-3.30) | (-2.45) |
| *Size*$_t$ | -0.229*** | -0.275*** | -0.293*** | -0.316*** | -0.339*** | -0.365*** | -0.374*** | -0.366*** |
| | (-5.89) | (-6.35) | (-6.66) | (-7.18) | (-8.14) | (-8.60) | (-8.87) | (-8.82) |
| Firm FE | Y | Y | Y | Y | Y | Y | Y | Y |
| YearQtr FE | Y | Y | Y | Y | Y | Y | Y | Y |
| R-squared | 0.744 | 0.719 | 0.711 | 0.706 | 0.702 | 0.694 | 0.688 | 0.688 |
| Observations | 66,705 | 65,794 | 64,914 | 64,015 | 63,110 | 62,278 | 61,426 | 60,535 |

**Table IA.4.** ChatGPT Investment Score and Future Returns: Controlling for Sentiment

This table reports coefficients from a firm-quarter level estimation that regresses firms' stock performance in the next quarter on the predicted capital expenditure by ChatGPT using earnings call transcripts. *ChatGPT Investment Score* measures the capital expenditure change predicted by ChatGPT from firms' earnings call transcripts. The dependent variable, *Stock Performance*, is one of the following three measures in quarter $t+2$: Annualized Quarterly Raw Return (*Return*); Annualized Quarterly Fama-French 5-factor alpha (*FF5-Adjusted Return*); Annualized Quarterly $q$-factor alpha (*q5-Adjusted Return*). Control variables include *Total q*, *Return*, and *Sentiment* in quarter $t$. All variables are defined in Appendix A. In all panels, the $t$-statistics, in parentheses, are based on standard errors clustered by firm. ***, **, * denote statistical significance at the 0.01, 0.05, and 0.10 levels, respectively.

| | (1) | (2) | (3) | (4) | (5) | (6) |
|---|---|---|---|---|---|---|
| | $Return_{t+2}$ | | $FF5\text{-}Adjusted\ Return_{t+2}$ | | $q5\text{-}Adjusted\ Return_{t+2}$ | |
| $ChatGPT\ Investment\ Score_t$ | -16.85*** | -9.319*** | -14.82*** | -7.358*** | -13.23*** | -6.582*** |
| | (-7.68) | (-4.19) | (-6.42) | (-3.15) | (-5.86) | (-2.88) |
| $Total\ q_t$ | | -15.67*** | | -13.13*** | | -12.70*** |
| | | (-19.54) | | (-15.85) | | (-14.97) |
| $Return_t$ | | -0.0597*** | | -0.155*** | | -0.0974*** |
| | | (-2.93) | | (-7.10) | | (-4.44) |
| $Sentiment_t$ | -5.472** | -3.372 | -7.305*** | -4.078 | -8.711*** | -6.322** |
| | (-2.15) | (-1.31) | (-2.74) | (-1.51) | (-3.20) | (-2.30) |
| Firm FE | Y | Y | Y | Y | Y | Y |
| YearQtr FE | Y | Y | Y | Y | Y | Y |
| R-squared | 0.233 | 0.239 | 0.087 | 0.094 | 0.083 | 0.088 |
| Observations | 74,429 | 74,429 | 74,429 | 74,429 | 74,429 | 74,429 |

**Table IA.5.** ChatGPT Investment Score and Future Return: Controlling for Factor Returns

This table presents coefficients from a firm-quarter level estimation that regresses return in quarter $t+2$ on the predicted capital expenditure by ChatGPT using earnings call transcripts, controlling for FF 5-factor returns or Q-5 factor returns. *ChatGPT Investment Score* measures the capital expenditure change predicted by ChatGPT from firms' earnings call transcripts. The dependent variables include *Total q* and the factor returns from the Fama-French 5-factor model (Fama and French, 2015) and the $q$5-factor model (Hou, Mo, Xue, and Zhang, 2021) in quarter $t$. All variables are defined in Appendix A. In all panels, the $t$-statistics, in parentheses, are based on standard errors clustered by firm. ***, **, * denote statistical significance at the 0.01, 0.05, and 0.10 levels, respectively.

| | (1) | (2) | (3) | (4) |
|---|---|---|---|---|
| | $Return_{t+2}$ | | | |
| *ChatGPT Investment Score$_t$* | -17.55*** | -10.22*** | -17.61*** | -10.27*** |
| | (-8.54) | (-4.90) | (-8.58) | (-4.93) |
| *Total $q_t$* | | -16.10*** | | -16.14*** |
| | | (-20.52) | | (-20.55) |
| *Ret_MKT$_t$* | 0.963*** | 0.964*** | | |
| | (40.84) | (40.96) | | |
| *Ret_Size$_t$* | 0.638*** | 0.640*** | | |
| | (12.34) | (12.53) | | |
| *Ret_IA$_t$* | 0.0323 | 0.0225 | | |
| | (0.57) | (0.40) | | |
| *Ret_ROE$_t$* | -0.216*** | -0.218*** | | |
| | (-5.93) | (-6.02) | | |
| *Ret_EG$_t$* | -0.108** | -0.0977* | | |
| | (-2.08) | (-1.90) | | |
| *Ret_MKT$_t$* | | | 0.990*** | 0.989*** |
| | | | (43.23) | (43.27) |
| *Ret_SMB$_t$* | | | 0.797*** | 0.799*** |
| | | | (18.71) | (18.70) |
| *Ret_HML$_t$* | | | 0.0262 | 0.0232 |
| | | | (0.78) | (0.70) |
| *Ret_RMW$_t$* | | | 0.107** | 0.110** |
| | | | (2.12) | (2.21) |
| *Ret_CMA$_t$* | | | -0.0672 | -0.0828 |
| | | | (-0.92) | (-1.15) |
| Firm FE | Y | Y | Y | Y |
| YearQtr FE | Y | Y | Y | Y |
| R-squared | 0.302 | 0.309 | 0.302 | 0.308 |
| Observations | 74,429 | 74,429 | 74,429 | 74,429 |

**Table IA.6.** Robustness Test: Out-of-Sample Test

This table presents coefficients from a firm-quarter level estimation that regresses firms' real capital expenditure in subsequent quarters on the predicted capital expenditure by ChatGPT with a subsample consisting of earnings calls occurring after the training period of ChatGPT 3.5 model from 2021Q4 to 2022Q4. *ChatGPT Investment Score* measures the capital expenditure change predicted by ChatGPT from firms' earnings call transcripts. The dependent variable, *Capital Expenditure*, is the real capital expenditure scaled by book assets for quarter $t+2$. Control variables include *Total q*, *Capital Expenditure*, *Total Cash Flow*, *Market Leverage*, and *Firm Size* in quarter $t$. All variables are defined in Appendix A. In all panels, the $t$-statistics, in parentheses, are based on standard errors clustered by firm. ***, **, * denote statistical significance at the 0.01, 0.05, and 0.10 levels, respectively.

| | (1) | (2) | (3) | (4) |
|---|---|---|---|---|
| | *Capital Expenditure*$_{t+2}$ | | | |
| *ChatGPT Investment Score*$_t$ | 1.152*** | 1.131*** | 0.729*** | 0.724*** |
| | (6.42) | (6.29) | (5.28) | (5.24) |
| *Total q*$_t$ | | 0.0125** | | 0.00592** |
| | | (2.47) | | (2.12) |
| *Capital Expenditure*$_t$ | | | 0.657*** | 0.656*** |
| | | | (28.02) | (27.99) |
| *Total Cash Flow*$_t$ | | | -0.00727 | -0.0150 |
| | | | (-0.36) | (-0.71) |
| *Leverage*$_t$ | | | -0.202*** | -0.175*** |
| | | | (-3.56) | (-2.99) |
| *Size*$_t$ | | | -0.000525 | -0.000891 |
| | | | (-0.09) | (-0.15) |
| Industry FE | Y | Y | Y | Y |
| YearQtr FE | Y | Y | Y | Y |
| R-squared | 0.185 | 0.186 | 0.506 | 0.506 |
| Observations | 10,470 | 10,470 | 10,470 | 10,470 |

**Table IA.7.** Robustness Test: Masked-Identity Test

This table presents coefficients from a firm-quarter level estimation that regresses firms' real capital expenditure in subsequent quarters on the predicted capital expenditure by ChatGPT, with a subsample where identifying information such as company, manager, and product names are removed from earnings call transcripts. *ChatGPT Investment Score* measures the capital expenditure change predicted by ChatGPT from firms' earnings call transcripts. The dependent variable, *Capital Expenditure*, is the real capital expenditure scaled by book assets for quarter $t+2$. Control variables include *Total q*, *Capital Expenditure*, *Total Cash Flow*, *Market Leverage*, and *Firm Size* in quarter $t$. All variables are defined in Appendix A. In all panels, the $t$-statistics, in parentheses, are based on standard errors clustered by firm. ***, **, * denote statistical significance at the 0.01, 0.05, and 0.10 levels, respectively.

| | (1) | (2) | (3) | (4) |
|---|---|---|---|---|
| | *Capital Expenditure*$_{t+2}$ | | | |
| *ChatGPT Investment Score*$_t$ | 0.417*** | 0.352*** | 0.235*** | 0.224*** |
| | (4.89) | (4.38) | (3.37) | (3.26) |
| *Total q*$_t$ | | 0.121*** | | 0.0412 |
| | | (3.96) | | (1.56) |
| *Capital Expenditure*$_t$ | | | 0.221*** | 0.221*** |
| | | | (7.02) | (7.03) |
| *Total Cash Flow*$_t$ | | | 0.586** | 0.482 |
| | | | (2.06) | (1.65) |
| *Leverage*$_t$ | | | -1.086*** | -1.018*** |
| | | | (-5.37) | (-4.91) |
| *Size*$_t$ | | | -0.0274 | -0.0303 |
| | | | (-0.56) | (-0.63) |
| Firm FE | Y | Y | Y | Y |
| YearQtr FE | Y | Y | Y | Y |
| R-squared | 0.614 | 0.618 | 0.650 | 0.650 |
| Observations | 7,259 | 7,259 | 7,259 | 7,259 |

**Table IA.8.** Robustness Test: Controlling for More Covariates

This table presents coefficients from a firm-quarter level estimation that regresses firms' real capital expenditure in subsequent quarters on the predicted capital expenditure by ChatGPT with more covariates. *ChatGPT Investment Score* measures the capital expenditure change predicted by ChatGPT from firms' earnings call transcripts. The dependent variable, *Capital Expenditure*, is the real capital expenditure scaled by book assets for quarter $t+2$. Control variables include *Total q*, calculated with updated market values at 0, 1 or 5 days after the earnings call, *Profitability*, *Sales Growth*, *Z-score*, more lags of *Capital Expenditure* and other controls in the baseline settings, including *Capital Expenditure*, *Total Cash Flow*, *Market Leverage*, and *Firm Size* in quarter $t$. All variables are defined in Appendix A. In all panels, the $t$-statistics, in parentheses, are based on standard errors clustered by firm. ***, **, * denote statistical significance at the 0.01, 0.05, and 0.10 levels, respectively.

| | (1) | (2) | (3) | (4) |
|---|---|---|---|---|
| | *Capital Expenditure*$_{t+2}$ | | | |
| *ChatGPT Investment Score*$_t$ | 0.289*** | 0.291*** | 0.291*** | 0.291*** |
| | (14.04) | (14.04) | (14.01) | (14.02) |
| *Total q*$_t$ | 0.0474*** | | | |
| | (5.97) | | | |
| *Total q*$_{t_c}$ | | 0.0237*** | | |
| | | (4.60) | | |
| *Total q*$_{t_c+1}$ | | | 0.0232*** | |
| | | | (4.49) | |
| *Total q*$_{t_c+5}$ | | | | 0.0227*** |
| | | | | (4.48) |
| *Profitability*$_t$ | 1.110*** | 1.119*** | 1.116*** | 1.115*** |
| | (6.15) | (6.22) | (6.19) | (6.19) |
| *Sales Growth*$_t$ | 1.818 | 1.764 | 1.716 | 1.700 |
| | (1.46) | (1.42) | (1.38) | (1.37) |
| *Z-score*$_t$ | -0.00867** | -0.00774** | -0.00769** | -0.00764** |
| | (-2.25) | (-2.00) | (-1.99) | (-1.98) |
| # of CapEx lags | 8 | 8 | 8 | 8 |
| Other Controls | Y | Y | Y | Y |
| Firm FE | Y | Y | Y | Y |
| YearQtr FE | Y | Y | Y | Y |
| R-squared | 0.749 | 0.748 | 0.748 | 0.748 |
| Observations | 63,699 | 63,699 | 63,699 | 63,699 |

**Table IA.9.** Robustness Test: Other Large Language Models

This table presents coefficients from a firm-quarter level estimation that regresses firms' real capital expenditure in subsequent quarters on the predicted capital expenditure by another large language model, RoBERTa. *ChatGPT Investment Score* measures the capital expenditure change predicted by ChatGPT from firms' earnings call transcripts. *RoBERTa Investment Score* measures the capital expenditure change predicted by RoBERTa from firms' earnings call transcripts. The dependent variable, *Capital Expenditure*, is the real capital expenditure scaled by book assets for quarter $t+2$. Control variables include *Total q*, *Capital Expenditure*, *Total Cash Flow*, *Market Leverage*, and *Firm Size* in quarter $t$. All variables are defined in Appendix A. In all panels, the $t$-statistics, in parentheses, are based on standard errors clustered by firm. ***, **, * denote statistical significance at the 0.01, 0.05, and 0.10 levels, respectively.

| | (1) | (2) | (3) |
|---|---|---|---|
| | *Capital Expenditure*$_{t+2}$ | | |
| *ChatGPT Investment Score*$_t$ | 0.477*** | | 0.444*** |
| | (13.91) | | (12.99) |
| *RoBERTa Investment Score*$_t$ | | 1.321*** | 0.980*** |
| | | (8.28) | (6.23) |
| *Total q*$_t$ | 0.108*** | 0.119*** | 0.107*** |
| | (7.47) | (8.06) | (7.34) |
| *Capital Expenditure*$_t$ | 0.149*** | 0.149*** | 0.149*** |
| | (8.91) | (8.91) | (8.92) |
| *Total Cash Flow*$_t$ | 0.607*** | 0.690*** | 0.579*** |
| | (3.54) | (4.04) | (3.38) |
| *Leverage*$_t$ | -1.302*** | -1.360*** | -1.286*** |
| | (-12.93) | (-13.23) | (-12.76) |
| *Size*$_t$ | -0.0974*** | -0.106*** | -0.102*** |
| | (-4.22) | (-4.58) | (-4.46) |
| Firm FE | Y | Y | Y |
| YearQtr FE | Y | Y | Y |
| R-squared | 0.575 | 0.573 | 0.575 |
| Observations | 74,429 | 74,429 | 74,429 |

**Table IA.10.** Robustness Test: Alternative Measure of ChatGPT Investment Score

This table presents coefficients from a firm-quarter level estimation that regresses firms' real capital expenditure in subsequent quarters on the predicted capital expenditure by ChatGPT using a different approach from Table 3. *ChatGPT Investment Alt. Score* assigns the ChatGPT-based text-chunk investment score with the largest absolute value to an earnings call. The dependent variable, *Capital Expenditure*, is the real capital expenditure scaled by book assets for quarter $t+2$. Control variables include *Total q*, *Capital Expenditure*, *Total Cash Flow*, *Market Leverage*, and *Firm Size* in quarter $t$. All variables are defined in Appendix A. In all panels, the $t$-statistics, in parentheses, are based on standard errors clustered by firm. ***, **, * denote statistical significance at the 0.01, 0.05, and 0.10 levels, respectively.

| | (1) | (2) | (3) | (4) |
|---|---|---|---|---|
| | *Capital Expenditure*$_{t+2}$ | | | |
| *ChatGPT Investment Alt. Score*$_t$ | 0.222*** | 0.196*** | 0.174*** | 0.167*** |
| | (12.34) | (11.31) | (10.76) | (10.41) |
| *Total q*$_t$ | | 0.243*** | | 0.119*** |
| | | (14.10) | | (8.09) |
| *Capital Expenditure*$_t$ | | | 0.149*** | 0.148*** |
| | | | (8.91) | (8.88) |
| *Total Cash Flow*$_t$ | | | 0.960*** | 0.711*** |
| | | | (5.67) | (4.15) |
| *Leverage*$_t$ | | | -1.542*** | -1.362*** |
| | | | (-14.92) | (-13.41) |
| *Size*$_t$ | | | -0.100*** | -0.102*** |
| | | | (-4.28) | (-4.38) |
| Firm FE | Y | Y | Y | Y |
| YearQtr FE | Y | Y | Y | Y |
| R-squared | 0.547 | 0.553 | 0.572 | 0.574 |
| Observations | 74,429 | 74,429 | 74,429 | 74,429 |

## Appendix D: Proofs

*Proof of Proposition 1.* First note that in period $t+1$, constant return to scale implies that

$$V_{t+1} = E_{t+1}[M_{t+1}\pi_{t+2}(K_{t+2})] = K_{t+2}q_{t+1}.$$

In the firm's investment problem in period $t+1$, the firm needs to optimize the following problem,

$$I_{t+1} = \text{argmax}_I \pi_{t+1}(K_{t+1}) - c(I, K_{t+1}) + ((1-\delta)K_{t+1} + I)q_{t+1}. \tag{6}$$

The first-order condition then implies that

$$c_1 + 2c_2 \frac{I_{t+1}}{K_{t+1}} = q_{t+1}. \tag{7}$$

Given that $q_{t+1} = q_t^e + q_t^m + \epsilon_{t+1}$, the expected value of $q_{t+1}$ increases with $q_m^t$. Therefore, investment $I_{t+1}$ is increasing in the managerial expectation $q_t^m$. □

*Proof of Proposition 2.* Since the pre-disclosure value $V_{t+1,d-}$ is independent of $q_t^m$, we only need to show that the post-disclosure value $V_{t+1,d+}$ is increasing in $q_t^m$. In fact,

$$V_{t+1,d+} = E_{t+1,d+}[\pi_{t+1}(K_{t+1}) - c(I^*_{t+1}, K_{t+1}) + ((1-\delta)K_{t+1} + I^*_{t+1})q_{t+1}]. \tag{8}$$

where $I^*_{t+1}$ is the solution to Equation (6). By the envelope theorem, the right hand side of (8) is increasing with the expected value of $q_{t+1}$. Therefore, the market value $V_{t+1,d+}$ is increasing in $q_t^m$. □

*Proof of Proposition 3.* We have

$$E_{t+1}[R_{t+1}] = \frac{E_{t+1}[V_{t+2}]}{V_{t+1}} = \frac{E_{t+1}[V_{t+2}]}{q_{t+1}K_{t+2}} = \frac{E_{t+1}[V_{t+2}/K_{t+2}]}{q_{t+1}}. \tag{9}$$

Therefore, other things equal, if the managerial expectations increases, $q_{t+1}$ decreases and the expected return in the next period declines. □